# Thermodynamics of amide + amine mixtures. 4. Relative permittivities of *N*,*N*-dimethylacetamide + *N*-propylpropan-1-amine, + *N*-butylbutan-1-amine, + butan-1-amine, or + hexan-1-amine systems and of *N*,*N*-dimethylformamide + aniline mixture at several temperatures. Characterization of amine + amide systems using ERAS

Fernando Hevia, Juan Antonio González*, Ana Cobos, Isaías García de la Fuente, Luis Felipe Sanz

G.E.T.E.F., Departamento de Física Aplicada, Facultad de Ciencias, Universidad de Valladolid, Paseo de Belén, 7, 47011 Valladolid, Spain.

*e-mail: jagl@termo.uva.es; Fax: +34-983-423136; Tel: +34-983-423757

**Abstract**

Relative permittivities at 1 MHz, $\varepsilon_{\mathrm{r}}$, and at (293.15-303.15) K are reported for the binary systems *N*,*N*-dimethylacetamide (DMA) + *N*-propylpropan-1-amine (DPA), + *N*-butylbutan-1-amine (DBA), + butan-1-amine (BA) or + hexan-1-amine (HxA) and for *N*,*N*-dimethylformamide (DMF) + aniline. The excess permittivities, $\varepsilon_{\mathrm{r}}^{\mathrm{E}}$ , are large and negative for systems with DMA, whereas they are large and positive for the aniline mixture. From the analysis of these $\varepsilon_{\mathrm{r}}^{\mathrm{E}}$ data and of measurements previously reported, it is concluded: (i) the main contribution to $\varepsilon_{\mathrm{r}}^{\mathrm{E}}$ in systems with linear amines arises from the breaking of interactions between like molecules; (ii) in the DMF + aniline mixture, interactions between unlike molecules contribute positively to $\varepsilon_{\mathrm{r}}^{\mathrm{E}}$, and such a contribution is dominant; (iii) longer linear amines are better breakers of the amide-amide interactions; (iv) interactions between unlike molecules are more easily formed when shorter linear amines, or DMF, participate. These findings are confirmed by a general study conducted in terms of excess values of molar orientational and induced polarizabilities and of the relative Kirkwood correlation factors for systems and components. The ERAS model is also applied to amide + amine mixtures. ERAS represents rather accurately the excess enthalpies and volumes of the mentioned systems. The variation of the cross-association equilibrium constants, determined using ERAS, with the molecular structure is in agreement with that observed for $\varepsilon_{\mathrm{r}}^{\mathrm{E}}$.

**Keywords**: Amides; amines; permittivity, Kirkwood correlation factor; excess functions; ERAS.

## 1. Introduction

The chemical environment of proteins is highly complex. A suitable approach for its investigation is to focus on small organic molecules which are more or less similar to the functional groups which constitute the biomolecule [1]. In this framework, the determination of thermodynamic, transport and dielectric properties for the mentioned molecules and for their mixtures is necessary, as information on interactions in condensed phase environments can be inferred from these properties.

Amides are a very important class of organic solvents due to their high polarity (the dipole moment of *N,N*-dimethylformamide (DMF) and *N,N*-dimethylacetamide (DMA) is 3.7 D [2,3]), strong solvating power and liquid state range [4]. The latter is strongly linked to the ability of amides to form hydrogen bonds. It is well known that primary and secondary amides are self-associated species, while tertiary amides show a relevant local order due to the existence of strong dipolar interactions between their molecules [5,6]. This makes amides useful as model systems for peptides [6].

The amine group is also encountered in substances of great biological interest. For example, histamine and dopamine act as neurotransmitters [7,8], and the breaking of amino acids releases amines. On the other hand the proteins usually bound to DNA polymers contain various amine groups [9]. Interestingly, primary and secondary amines are self-associated compounds [10-14] with low dipole moments in the case of linear amines (1.3 D for BA and 1.0 D for DPA [15]). The dipole moment of aniline (1.51 D [3]) is higher and proximity effects between the phenyl ring and the amine group lead to strong dipolar interactions between aniline molecules. As a consequence, aniline + *n*-alkane mixtures are characterized by relatively high upper critical solution temperatures (343.1 K for the heptane solution [16]).

The study of amine + amide systems is relevant as it allows to gain insight into the amide group behaviour when it is surrounded by different environments. In fact, the hydrogen-bonded structures where the amide group is involved can show very different biological activities depending on the mentioned environments [17].

The few data available in the literature on excess molar enthalpies, $H_{\mathrm{m}}^{\mathrm{E}}$, for amine + amide mixtures underline the importance of interactions between unlike molecules in such systems. For example, at equimolar composition, we have $H_{\mathrm{m}}^{\mathrm{E}}/\mathrm{J\cdot mol^{-1}} = -2946$ (aniline + DMF, $T$ = 298.15 K) [18]; $-352$ (aniline + DMA, $T$ = 298.15 K) [19], $-1000$ (HxA + *N*-methylacetamide (NMA), $T$ = 363.15 K) [20]. Interestingly, $H_{\mathrm{m}}^{\mathrm{E}}/\mathrm{J\cdot mol^{-1}}$ values of methanol + NMA ($-76$, $T$ = 313.15 K) [21], or + DMA ($-737$; $T$ = 298.15 K) [22] are very different.

In previous studies, we have reported data on density, $\rho$, speed of sound, *c*, and refractive index, $n_D$, for the binary systems DMF [23], or DMA [24] + *N*-propylpropan-1-amine (DPA) or + butan-1-amine (BA) at (293.15-303.15) K, and + *N*-butylbutan-1-amine (DBA) or + hexan-1-amine (HxA) at 298.15 K. These data have been interpreted in terms of solute-solvent interactions and structural effects [23,24]. On the other hand, we have also reported permittivity measurements for the DMF + BA, + HxA, + DPA, + DBA systems at (293.15-303.15) K [25]. As a continuation of these works, we provide now low-frequency relative permittivities, $\varepsilon_r$, for the DMA + BA, + HxA, + DPA, + DBA mixtures, and for the DMF + aniline system at the same temperature range. The replacement of DMF by DMA in the mentioned systems including linear amines may be useful to investigate steric/size effects on the excess $\varepsilon_r$ values. The aniline + DMF system has been selected on the basis of its very large and negative $H_m^E$ value. The present study is completed by the application of different theories. Firstly, amine + amide mixtures are studied using the ERAS model [26]. Secondly, the $\varepsilon_r$ data reported here are used together with the corresponding $\rho$ and $n_D$ values available in the literature [23,24,27] to determine orientational and induced polarizabilities according to the Kirkwood-Fröhlich model [28-31] and the Balankina relative excess Kirkwood correlation factors [32], very useful quantities to gain insight into the dipole correlations present in the mixtures under consideration.

## 2. Experimental

### *2.1 Materials*

Table 1 collects information regarding the source and purity of the pure compounds, which have been used with no further purification

### *2.2 Apparatus and procedure*

Binary mixtures were prepared by mass in small vessels of about 10 $cm^3$, using an analytical balance Sartorius MSU125p (weighing accuracy 0.01 mg), with all weighings corrected for buoyancy effects. The standard uncertainty in the final mole fraction is estimated to be 0.0010. Molar quantities were calculated using the relative atomic mass Table of 2015 issued by the Commission on Isotopic Abundances and Atomic Weights (IUPAC) [33]. In order to minimize the effects of the interaction of amines with air components, they were stored with 4 Å molecular sieves; also, the measurement cell (see below) was completely filled with the samples and appropriately closed. Different density measurements of pure compounds, conducted along experiments, showed that this quantity remained unchanged within the experimental uncertainty.

Temperatures were measured by means of Pt-100 resistances, calibrated according to the ITS-90 scale of temperature, against two fixed points: the triple point of water and the fusion point of Ga. The standard uncertainty of the equilibrium temperature measurements is 0.01 K and the corresponding accuracy is 0.02 K.

Permittivity measurements were conducted using a 16452A cell (parallel-plate capacitor) connected, by means of a 16048G test lead, to a precision impedance analyser 4294A; all of them are from Agilent. The 16452A cell is made of Nickel-plated cobalt (54% Fe, 17% Co, 29% Ni) with a ceramic insulator (alumina, $Al_2O_3$). The volume of the sample filling the cell is $\approx 4.8$ cm$^3$. The temperature was controlled by a thermostatic bath LAUDA RE304, (temperature stability: 0.02 K). Details about the device configuration and calibration can be found elsewhere [34]. The relative standard uncertainty of the $\varepsilon_r$ measurements (i.e. the repeatability) is 0.0001. The total relative standard uncertainty of $\varepsilon_r$ was estimated to be 0.003 from the differences between our data and values available in the literature for the following pure liquids in the temperature range (288.15–333.15) K: water, benzene, cyclohexane, hexane, nonane, decane, dimethyl carbonate, diethyl carbonate, methanol, 1-propanol, 1-pentanol, 1-hexanol, 1-heptanol, 1-octanol, 1-nonanol and 1-decanol.

Our experimental $\varepsilon_\mathrm{r}$ values, at 1 MHz and 0.1 MPa, of pure compounds, together with literature data, are shown in Table 2. We note the excellent agreement encountered between them for DMF and DMA. Larger discrepancies between such data are observed for amines, which may be ascribed to the different source and purity of the amines used in the literature. In fact, inspection of Table 2 shows that, for example, some $\varepsilon_r$ values of aniline taken from the literature are not sure as they do not change consistently with temperature. In contrast, our $\varepsilon_\mathrm{r}$ values correctly decrease with the increasing of temperature, and the density measurements are in good agreement with literature data (Table S1, supplementary material; see also [23,24] for the remaining amines).

## 3. Experimental results

The relative permittivity of an ideal mixture at the same temperature and pressure as the solution under study, $\varepsilon_\mathrm{r}^\mathrm{id}$, is calculated from the expression [35]:

$$\varepsilon_\mathrm{r}^\mathrm{id} = \phi_1 \varepsilon_\mathrm{r1}^* + \phi_2 \varepsilon_\mathrm{r2}^* \tag{1}$$

where the volume fraction of component $i$ is defined as $\phi_i = x_i V_{\mathrm{m}i}^* / V_\mathrm{m}^\mathrm{id}$; $x_i$ represents the mole fraction of component $i$, $V_{\mathrm{m}i}^*$ and $\varepsilon_{\mathrm{r}i}^*$ stand for the molar volume and relative permittivity of pure component $i$ respectively, and $V_{\mathrm{m}}^{\mathrm{id}} = x_1 V_{\mathrm{m1}}^{*} + x_2 V_{\mathrm{m2}}^{*}$ is the ideal molar volume of the mixture at the same temperature and pressure. The excess relative permittivity, $\varepsilon_{\mathrm{r}}^{\mathrm{E}}$, is obtained as

$$\varepsilon_{\mathrm{r}}^{\mathrm{E}} = \varepsilon_{\mathrm{r}} - \varepsilon_{\mathrm{r}}^{\mathrm{id}} \qquad (2)$$

where $\varepsilon_{\mathrm{r}}$ is the permittivity of the mixture. The necessary volumetric properties were obtained from the literature [23,24,27] (see also footnote of Table 2). Table 3 lists $\phi_1$, $\varepsilon_{\mathrm{r}}$ and $\varepsilon_{\mathrm{r}}^{\mathrm{E}}$ values for DMA (1) + amine (2), or DMF (1) + aniline (2) systems as functions of the mole fraction of the amide, $x_1$, in the temperature range (293.15 – 303.15) K. Results are shown graphically in Figures 1-3 (see also Figure S1, supplementary material). The only data available in the literature [36] for comparison are those for the DMF + aniline system at 303.15 K. They largely differ from our measurements (Figure S2, supplementary material).

The $\varepsilon_{\mathrm{r}}^{\mathrm{E}}$ data have been fitted by an unweighted linear least-squares regression to a Redlich-Kister equation:

$$\varepsilon_{\mathrm{r}}^{\mathrm{E}} = x_1(1-x_1)\sum_{i=0}^{k-1} A_i (2x_1 - 1)^i \qquad (3)$$

For each system, the number, $k$, of necessary coefficients for this regression has been determined by applying an F-test of additional term [37] at 99.5% confidence level. Table 4 includes the parameters $A_i$ obtained, and the standard deviations $\sigma(\varepsilon_{\mathrm{r}}^{\mathrm{E}})$, defined by:

$$\sigma(\varepsilon_{\mathrm{r}}^{\mathrm{E}}) = \left[\frac{1}{N-k}\sum_{j=1}^{N}\left(\varepsilon_{\mathrm{r,cal},j}^{\mathrm{E}} - \varepsilon_{\mathrm{r,exp},j}^{\mathrm{E}}\right)^2\right]^{1/2} \qquad (4)$$

where $N$ is the number of $\varepsilon_{\mathrm{r,exp},j}^{\mathrm{E}}$ experimental data, and $\varepsilon_{\mathrm{r,cal},j}^{\mathrm{E}}$ is the corresponding value of the excess property $\varepsilon_{\mathrm{r}}^{\mathrm{E}}$ calculated from equation (3).

## 4. ERAS model

Some important features of this model are now given. (i) The excess functions are calculated as the sum of two contributions. The chemical contribution, $F_{\mathrm{m,chem}}^{\mathrm{E}}$, arises from hydrogen-bonding; the physical contribution, $F_{\mathrm{m,phys}}^{\mathrm{E}}$, is related to non-polar Van der Waals' interactions including free volume effects. Expressions for the molar excess functions $F_{\mathrm{m}}^{\mathrm{E}} = H_{\mathrm{m}}^{\mathrm{E}}$ (enthalpy); $V_{\mathrm{m}}^{\mathrm{E}}$ (volume) can be found elsewhere [38,39]. (ii) It is assumed that only consecutive linear association occurs. Such an association is described by a chemical equilibrium constant ($K_{\mathrm{A}}$) independent of the chain length of the associated species (amines), according to the equation:

$$\mathrm{A}_m + \mathrm{A} \leftrightarrow \mathrm{A}_{m+1} \quad (5)$$

with $m$ ranging from 1 to $\infty$. The cross-association between a self-associated species $A_m$ and a non self-associated compound $B$ (in this study, tertiary amides) is represented by

$$A_m + B \xleftrightarrow{K_{\mathrm{AB}}} A_m B \quad (6)$$

Linear secondary amides (*N*-methylacetamide is also considered in this work) are also self-associated and their association is described by an equation similar to equation (5):

$$B_n + B \leftrightarrow B_{n+1} \quad (7)$$

with $n$ ranging from 1 to $\infty$. The cross-association is then represented by:

$$A_m + B_n \xleftrightarrow{K_{\mathrm{AB}}} A_m B_n \quad (8)$$

The cross-association constants ($K_{\mathrm{AB}}$) of equations (6) and (8) are also considered to be independent of the chain length. Equations (5)-(8) are characterized by $\Delta h_{\mathrm{i}}^*$, the enthalpy of the reaction that corresponds to the hydrogen-bonding energy, and by the volume change ($\Delta v_{\mathrm{i}}^*$) related to the formation of the linear chains. (iii) The $F_{\mathrm{m,phys}}^{\mathrm{E}}$ term is derived from the Flory's equation of state [40], which is assumed to be valid not only for pure compounds but also for the mixture [41,42]:

$$\frac{\bar{P}_{\mathrm{i}}\bar{V}_{\mathrm{i}}}{\bar{T}_{\mathrm{i}}} = \frac{\bar{V}_{\mathrm{i}}^{1/3}}{\bar{V}_{\mathrm{i}}^{1/3} - 1} - \frac{1}{\bar{V}_{\mathrm{i}}\,\bar{T}_{\mathrm{i}}} \quad (9)$$

where i = A,B or M (mixture). In equation (9), $\bar{V}_{\mathrm{i}} = V_{\mathrm{mi}} / V_{\mathrm{i}}^*$; $\bar{P}_{\mathrm{i}} = P / P_{\mathrm{i}}^*$; $\bar{T}_{\mathrm{i}} = T / T_{\mathrm{i}}^*$ are the reduced properties for volume, pressure and temperature, respectively. The pure component reduction parameters $V_{\mathrm{i}}^*$, $P_{\mathrm{i}}^*$, $T_{\mathrm{i}}^*$ are obtained from *P-V-T* data (density, $\alpha_{\mathrm{p}}$, isobaric thermal expansion coefficient, and isothermal compressibility, $\kappa_T$), and association parameters [41,42]. The reduction parameters for the mixture $P_{\mathrm{M}}^*$ and $T_{\mathrm{M}}^*$ are calculated from mixing rules [41,42]. The total relative molecular volumes and surfaces of the compounds were calculated additively on the basis of the group volumes and surfaces recommended by Bondi [43].

### *4.1 Adjustment of ERAS parameters*

Values of $V_{\mathrm{i}}$, $V_{\mathrm{i}}^*$ and $P_{\mathrm{i}}^*$ of pure compounds at $T$ = 298.15 K, needed for calculations, are listed in Table S1 of supplementary material. $K_{\mathrm{A}}$, $\Delta h_{\mathrm{A}}^*$, and $\Delta v_{\mathrm{A}}^*$ of the self-associated amines and of *N*-methylacetamide are known from $H_{\mathrm{m}}^{\mathrm{E}}$ and $V_{\mathrm{m}}^{\mathrm{E}}$ data for the corresponding mixtures with alkanes [11-13,44]. The binary parameters to be fitted against $H_{\rm m}^{\rm E}$ [18-20] and $V_{\rm m}^{\rm E}$ [23,24,27,45] data available in the literature for amine + amide systems are then $K_{\rm AB}$, $\Delta h_{\rm AB}^{*}$, $\Delta v_{\rm AB}^{*}$ and $X_{\rm AB}$. They are collected in Table 5.

### 4.2 Results

ERAS results are shown in Table 6 and Figures 4 and 5 (see also Figures S3 and S4 of supplementary material). We must underline that the model describes, rather correctly, the $H_{\rm m}^{\rm E}$ and $V_{\rm m}^{\rm E}$ functions of the amine + amide systems under study using parameters which smoothly change with the molecular structure. (Table 5).

## 5. Kirkwood-Fröhlich model

Some relevant hypotheses of the model are: (i) molecules of a given polar compound are assumed to be spherical (i.e., an intrinsic dipole moment inside a spherical cavity), (ii) the effect of the induced polarization of the molecules is treated in macroscopic way, assuming that the cavity is filled by a continuous medium of relative permittivity $\varepsilon_{\rm r}^{\infty}$ (the value of the permittivity at a high frequency at which only the induced polarizability contributes); (iii) long-range interactions are taken into account macroscopically by considering the outside of the cavity as a continuous dielectric of permittivity $\varepsilon_{\rm r}$, leading to the Onsager local field; (iv) short-range interactions are not neglected. A central magnitude of the theory is the so-called Kirkwood correlation factor, $g_{\rm K}$, which provides information of the deviations from randomness of the orientation of a dipole with respect to its neighbours. This is an important parameter, as it provides information on specific interactions in the liquid state. For a mixture, $g_{\rm K}$ can be determined, in the context of a one-fluid model [32], from macroscopic physical properties according to the expression [28,29,31,32]:

$$g_{\rm K} = \frac{9k_{\rm B}TV_{\rm m}\varepsilon_0(\varepsilon_{\rm r}-\varepsilon_{\rm r}^{\infty})(2\varepsilon_{\rm r}+\varepsilon_{\rm r}^{\infty})}{N_{\rm A}\mu^2\varepsilon_{\rm r}(\varepsilon_{\rm r}^{\infty}+2)^2} \quad (10)$$

Here, $k_{\rm B}$ is Boltzmann's constant; $N_{\rm A}$, Avogadro's constant; $\varepsilon_0$, the vacuum permittivity; and $V_{\rm m}$, the molar volume of the liquid at the working temperature, *T*. For polar compounds, $\varepsilon_{\rm r}^{\infty}$ is estimated from the relation $\varepsilon_{\rm r}^{\infty} = 1.1n_{\rm D}^2$ [46]. $\mu$ represents the gas phase dipole moment of the solution, estimated from the equation [32]:

$$\mu^2 = x_1\mu_1^2 + x_2\mu_2^2 \quad (11)$$

where $\mu_\mathrm{i}$ stands for the dipole moment of component i (=1,2) (Table 2).

The molar orientational polarizabilities (molar orientational polarizations or molar polarizability volumes), $\Pi_\mathrm{m}^\mathrm{or}$, are determined from [29-31]:

$$\Pi_\mathrm{m}^\mathrm{or} = \frac{N_A \alpha_\mathrm{or}}{3\varepsilon_0} = \frac{(\varepsilon_\mathrm{r} - \varepsilon_\mathrm{r}^\infty)(2\varepsilon_\mathrm{r} + \varepsilon_\mathrm{r}^\infty)}{9\varepsilon_\mathrm{r}} V_\mathrm{m} \qquad (12)$$

where $\alpha_\mathrm{or}$ stands for the orientational polarizability (in the case of mixtures, a one-fluid approach is implicit).

Molar induced polarizabilities, $\Pi_\mathrm{m}^\mathrm{ind}$, can be calculated in the framework of the Kirkwood-Fröhlich model by means of the expression:

$$\Pi_\mathrm{m}^\mathrm{ind} = \frac{N_A \alpha_\mathrm{ind}}{3\varepsilon_0} = \frac{(\varepsilon_\mathrm{r}^\infty - 1)(2\varepsilon_\mathrm{r} + \varepsilon_\mathrm{r}^\infty)}{9\varepsilon_\mathrm{r}} V_\mathrm{m} \qquad (13)$$

with $\alpha_\mathrm{ind}$ meaning the induced polarizability. Excess values of $\Pi_\mathrm{m}^\mathrm{or}$ and $\Pi_\mathrm{m}^\mathrm{ind}$ (Table 7, Figures 6 and S5, and S6 of supplementary material) have been obtained from the equation:

$$F^\mathrm{E} = F - F^\mathrm{id} \qquad (F = \Pi_\mathrm{m}^\mathrm{or} \text{ or } \Pi_\mathrm{m}^\mathrm{ind}) \qquad (14)$$

with $F^\mathrm{id}$ values determined from equations (12) or (13) using ideal values for the involved quantities in the mentioned equations. Particularly, calculations have been conducted using smoothed values of $V_\mathrm{m}^\mathrm{E}$ [23,24,27,45], $n_\mathrm{D}^\mathrm{E}$ [23,24,27] and $\varepsilon_\mathrm{r}^\mathrm{E}$ (this work) at $\Delta x_1$ = 0.01.

## 6. Discussion

Along the present section, the values of the physical properties which involve some permittivity measurements and of their excess functions are referred to 298.15 K and $\phi_1 = 0.5$. Values of $H_\mathrm{m}^\mathrm{E}$ and $V_\mathrm{m}^\mathrm{E}$ are referred to 298.15 K and equimolar composition.

It is known that the disruption of interactions between like molecules, in the present case amide-amide and amine-amine interactions, contributes negatively to $\varepsilon_\mathrm{r}^\mathrm{E}$. For instance, the $\varepsilon_\mathrm{r}^\mathrm{E}$ values of *n*-alkylamine + *n*-$C_{12}$ systems at 293.15 K are: $-0.314$ (propylamine) < $-0.243$ (BA) < $-0.133$ (HxA) [47]. This negative contribution diminishes when increasing the chain length of the amine, as the amine group is then more sterically hindered, in such a way that the effective polarity of longer amines becomes weaker. The creation of interactions between unlike molecules along the mixing process may lead either to a positive or to a negative contribution to $\varepsilon_\mathrm{r}^\mathrm{E}$ [48]. A positive contribution is encountered when interactions between unlike molecules lead to an increased number of effective dipole moments in the system. Negative contributions arise when interactions between unlike molecules lead to a loss of the polar structure of the liquid, and therefore to a decreased number of effective dipole moments.

The large and negative $\varepsilon_r^E$ values of DMA + linear amine mixtures reveal that the negative contributions from the breaking of interactions between like molecules are dominant. It is noteworthy that the $\varepsilon_r^E$ values of *n*-alkylamine + *n*-$C_{12}$ systems at 293.15 K are much less negative than those of DMA + *n*-alkylamine, e.g, $-2.447$ for the DPA system (see above). This suggests that $\varepsilon_r^E$ of DMA solutions is determined, to a large extent, by the breaking of the dipolar interactions between DMA molecules. On the other hand, one can expect that interactions between unlike molecules contribute positively to $\varepsilon_r^E$. In fact, the $\varepsilon_r^E$ value of the DMF + heptane mixture at $\phi_1 = 0.0171$ and 293.15 K is lower (– 0.24, calculated from data of the literature [49]) than the values of the corresponding systems with amines at the same conditions: – 0.129 (DPA), – 0.146 (DBA), – 0.104 (BA), and – 0.137 (HxA) [47]. Interestingly, $\varepsilon_r^E$ is positive for the DMF + aniline mixture. This clearly indicates that $\varepsilon_r^E$ is now mainly determined by the positive contribution related to the aniline-DMF interactions created upon mixing. Other systems as methanol + DMF (2.57 [50]); + DMA (0.52 [51]); + pyridine (2.85 [50]), or + cyclohexylamine (1.13 [52]) also show positive $\varepsilon_r^E$ values.

We note that, in DMA solutions, $\varepsilon_r^E$(DBA) < $\varepsilon_r^E$(DPA) and $\varepsilon_r^E$(HxA) < $\varepsilon_r^E$(BA) (Table 7, Figures 1,2). This can be explained as follows: (i) longer amines are better breakers of the DMA-DMA interactions due to their large aliphatic surface; (ii) the formation of interactions between unlike molecules becomes easier when shorter amines are involved, as the amine group is then less sterically hindered. It is remarkable that $\varepsilon_r^E$(HxA) ≈ $\varepsilon_r^E$(DPA), which suggests that the $\varepsilon_r$ decrease when HxA is replaced by DPA (note that $\varepsilon_r^*$(HxA) = 3.893 > $\varepsilon_r^*$(DPA) = 3.093, Table 2) is compensated by the creation of a larger number of interactions between unlike molecules in the case of the HxA mixture. Similar trends are observed for the corresponding systems with DMF, but an interesting difference is that $\varepsilon_r^E$(HxA) = $-1.383$ > $\varepsilon_r^E$(DPA) = $-1.509$ [25]. It seems that interactions between unlike molecules could be now even more important, as the amide group is less sterically hindered in DMF. Comparison between $\varepsilon_r^E$ values of mixtures with a given linear amine shows that $\varepsilon_r^E$(DMF) > $\varepsilon_r^E$(DMA). In addition, $\varepsilon_r^E$ curves of the DMA systems are more skewed towards larger $\phi_1$ values (Table 7). This suggests that linear amines can disrupt more easily DMA-DMA interactions and that the creation of amide-amine interactions is favoured when DMF molecules participate.

Finally, we must remark that the replacement of HxA ($\varepsilon_r^E$ = – 1.383) by aniline ($\varepsilon_r^E$ = 1.806) in DMF solutions has a large impact on the $\varepsilon_r^E$ values of these mixtures, which show opposite signs. Therefore, the aromaticity effect leads here to an increase of the number of effective dipole moments in the aniline system.

### *6.1 Temperature dependence of $\varepsilon_r$*

Some important information on interactions and structure of a liquid can be inferred from the temperature dependence of $\varepsilon_r$. Of particular interest is the investigation of entropic effects induced in the liquid by the external electric field applied, $\vec{E}$. The relationship between $(\partial\varepsilon_r / \partial T)$ and the entropy increment per volume unit is given by [28,53-55]:

$$\frac{\Delta S}{\vec{E}^2} = \frac{S(T,\vec{E}) - S_0(T)}{\vec{E}^2} = \frac{\varepsilon_0}{2}\left(\frac{\partial\varepsilon_r}{\partial T}\right) \qquad (15)$$

In this expression, $S(T,\vec{E})$ is the entropy per volume unit of the system at temperature $T$ under the application of $\vec{E}$, and $S_0(T)$ is the entropy per volume unit of the solution in absence of $\vec{E}$. The data analysis is more properly conducted on the basis of the $\frac{\Delta S}{\vec{E}^2}V_m$ magnitude as then one is considering along the discussion a number of molecules equal to $N_A$ [54,55]. Within this treatment, volume variations with $T$ have been neglected. We must note that $\Delta S$ < 0 corresponds to the dipolar ordering action of $\vec{E}$, which is the normal behaviour of common liquids. All the pure compounds and systems along the present work follow this trend (Tables 2 and 8). From our results, some conclusions can be stated. (i) The molar entropy increments induced in amides are much more negative than those induced in amines (Table 2). This remarks the existence of strong dipolar interactions between amide molecules, which lead to the formation of entities of high polarity. Such entities are better oriented by the application of $\vec{E}$. (ii) For linear amines, the molar entropy increments become less negative in the sequence: BA > HxA > DPA > DBA (Table 2). Clearly, it can be ascribed to a meaningful decrease of the orientational polarizability of the amines in the same order (see below). Aniline shows the largest $\left|\frac{\Delta S}{\vec{E}^2}V_m\right|$ value, as a consequence of its stronger polar character. (iii) Interestingly, $\frac{\Delta S}{\vec{E}^2}V_m$ is more negative for DMA than for DMF (Table 2). This suggests that the ability of the DMA entities to respond to the ordering action of $\vec{E}$ is higher, a behavior that can be attributed to weaker dipolar interactions between DMA molecules. This is supported experimentally by liquid-liquid equilibrium measurements, as the upper critical solution temperatures of heptane systems are 342.55 K (DMF) [56] > 309.8 K (DMA) [57] (see also the results from the Kirkwood-Fröhlich model below). (iv) A similar analysis is still valid for the considered mixtures (Table 8). For example, $\frac{\Delta S}{\vec{E}^2}V_\mathrm{m}$ values of DMA solutions decrease in the sequence BA > DBA. This can be explained assuming, as previously, that DBA is a better breaker of interactions between amide molecules leading to a higher loss of the polar structure of DMA. This makes that the remaining DMA entities, of smaller size than those in pure amide, can be more easily oriented by the action of $\vec{E}$ . It must be also remarked that the higher polar structure of the DMF + aniline system compared to that of the HxA system leads to a lower $\frac{\Delta S}{\vec{E}^2}V_\mathrm{m}$ value for the former solution.

## 6.2 *Results from ERAS model*

The application of the ERAS model is useful to complete the description given above. Some important features are now given. (i) Amine-amide interactions are rather strong, as $\Delta h^*_\mathrm{AB} = -22$ kJ·mol$^{-1}$, is a value not far from that used for 1-alkanol self-association ( – 25.1 kJ·mol$^{-1}$) in applications of the ERAS model [10,26,39,58]. It must be remarked that the same $\Delta h^*_\mathrm{AB}$ parameter is valid for all the tertiary amide + amine mixtures under consideration. (ii) Negative $V^\mathrm{E}_\mathrm{m}$ values of BA, or DPA + DMA system arise from structural effects, as it is suggested by $V^\mathrm{E}_\mathrm{m,phys}$ < 0 (Table 6) and positive $X_\mathrm{AB}$ values (Table 5). The $V^\mathrm{E}_\mathrm{m,chem}$ term (i.e., interactions between unlike molecules) is also relevant for the BA or DPA + DMF mixtures. Structural effects contribute more largely to $V^\mathrm{E}_\mathrm{m}$ in DPA systems. (iii) The large $\left|V^\mathrm{E}_\mathrm{m,chem}\right|$ values of aniline mixtures may be indicative that the model overestimates this contribution. In fact, the $V^\mathrm{E}_\mathrm{m}$ values of such systems are rather similar ($V^\mathrm{E}_\mathrm{m}$ /cm$^3$·mol$^{-1}$ = – 0.662 (DMF) [27]; – 0.636 (DMA) [45], $T$ = 303.15 K), while the corresponding $H^\mathrm{E}_\mathrm{m}$ values are very different (see above); (iv) The equilibrium constant $K_\mathrm{AB}$ decreases in the sequences: BA > HxA and DPA > DBA. In addition, for mixtures with a given amine, $K_\mathrm{AB}$ (DMF) > $K_\mathrm{AB}$ (DMA). If one takes into account that, in the ERAS model, self-association or solvation effects are described by means of linear chains formed by the system components, the relative variation of $K_\mathrm{AB}$ agrees with that encountered for $\varepsilon^\mathrm{E}_\mathrm{r}$ . On the other hand, the large $K_\mathrm{AB}$ value for the DMF + aniline mixture is to be noted, as it remarks that interactions between unlike molecules become here rather important.

## 6.3 *Results from Kirkwood-Fröhlich's theory*

Firstly, we give values of $(\Pi_m^{or})^*$ and of $(\Pi_m^{ind})^*$ for pure compounds: $(\Pi_m^{or})^*$/ cm$^3$·mol$^{-1}$ = 623.0 (DMF); 773.0 (DMA); 68.0 (BA); 64.1 (HxA); 38.6 (DPA); 36.3 (DBA), 103.0 (aniline) and $(\Pi_m^{ind})^*$/cm$^3$·mol$^{-1}$ = 22.0 (DMF); 27.0 (DMA); 31.4 (BA); 45.7 (HXA); 47.9 (DPA); 63.0 (DBA), 42.7 (aniline). It is remarkable that $(\Pi_m^{or})^*$ of DMA is much larger than $(\Pi_m^{ind})^*$. It is also important to consider the sum of these two quantities (i.e. the total molar polarizability); in the same units: 645.0 (DMF); 800.0 (DMA); 99.4 (BA); 109.8 (HxA); 86.5 (DPA); 99.3 (DBA). These results indicate that DMA molecules are more easily oriented by the application of an electric field than DMF molecules, and underline that dipolar interactions between DMA molecules are weaker than in DMF. This was suggested by their $\frac{\Delta S}{\vec{E}^2}V_m$ values (see above). Interestingly, the total molar polarizability of the compounds does not vary in the same sense as $\varepsilon_r^*$. Nevertheless, it must be taken into account that, when applying a certain electric field, $(\varepsilon_r^* - 1)$ is proportional to the macroscopic dipole moment per unit volume, and therefore the trend of $\varepsilon_r^*$ can be explained as arising from volume effects. In fact, the total molar polarizability per unit volume does vary accordingly to $\varepsilon_r^*$: 8.331 (DMF); 8.600 (DMA); 0.995 (BA); 0.825 (HxA); 0.627 (DPA); 0.580 (DBA), 1.54 (aniline).

Results for $(\Pi_m^{or})^E$ /(in cm$^3$·mol$^{-1}$) of DMA systems are: $-42.6$ (BA) > $-56.4$ (HxA) > $-59.2$ (DPA) > $-69.5$ (DBA). These $(\Pi_m^{or})^E$ values change in line with those of $\varepsilon_r^E$ (Table 7), pointing out to a main contribution to $\varepsilon_r^E$ arising from effects related to the orientational polarizability of the molecules. The same trend is observed for $(\Pi_m^{or})^E$ /cm$^3$·mol$^{-1}$ of DMF mixtures: 30.5 (aniline) > – 18.0 (BA) > – 27.8 (HxA) > – 31.6 (DPA) > – 41.4 (DBA). It seems to be clear that there is a loss of effective dipole moments in DMA + linear amines mixtures with regards to those involving DMF. In contrast, there is a meaningful increment of the effective dipole moments when HxA is replaced by aniline in DMF solutions. Interestingly, the $(\Pi_m^{ind})^E$ curves (Figures S5 and S6 of supplementary material) of *n*-alkylamine systems show a maximum at the concentrations where a minimum exists for the $(\Pi_m^{or})^E$ curves (Figure 6). This is a consistent result, as it indicates that the decrease of orientational polarization effects is linked to an increase of $\Pi_m^{ind}$, that is, roughly speaking, to an increase of dispersive interactions. For a given linear amine, the $(\Pi_m^{ind})^E$ maximum changes in the order: DMF < DMA, which also supports the more polar character of DMF solutions (Figures S5 and S6). In systems with a fixed amide, the mentioned maximum changes in the sequence: DBA > DPA > HxA > BA. That is, it decreases when the polarity of the mixture increases. It is to be noted that the increase of polarity in the DMF + aniline system is accompanied by a decrease in the dispersive interactions (Figures 6 and S6, supplementary material).

The Balankina relative excess Kirkwood correlation factors [32], $g_{\mathrm{K,rel}}^{\mathrm{E}} = (g_{\mathrm{K}} - g_{\mathrm{K}}^{\mathrm{id}}) / g_{\mathrm{K}}^{\mathrm{id}}$, where $g_{\mathrm{K}}$ and $g_{\mathrm{K}}^{\mathrm{id}}$ account respectively for the real and ideal Kirkwood correlation factors, are a useful tool to probe into the structure of the mixtures:

$$g_{\mathrm{K,rel}}^{\mathrm{E}} = \frac{V_{\mathrm{m}}(\varepsilon_{\mathrm{r}} - \varepsilon_{\mathrm{r}}^{\infty})(2\varepsilon_{\mathrm{r}} + \varepsilon_{\mathrm{r}}^{\infty})\varepsilon_{\mathrm{r}}^{\mathrm{id}}(\varepsilon_{\mathrm{r}}^{\mathrm{id},\infty} + 2)^2}{V_{\mathrm{m}}^{\mathrm{id}}(\varepsilon_{\mathrm{r}}^{\mathrm{id}} - \varepsilon_{\mathrm{r}}^{\mathrm{id},\infty})(2\varepsilon_{\mathrm{r}}^{\mathrm{id}} + \varepsilon_{\mathrm{r}}^{\mathrm{id},\infty})\varepsilon_{\mathrm{r}}(\varepsilon_{\mathrm{r}}^{\infty} + 2)^2} - 1 \qquad (16)$$

It is also possible to develop a two-liquid model [32], in which liquid i (i=1,2) is defined by molecules of pure substance i located in spherical cavities of volume $\bar{V}_{\mathrm{mi}} / N_A$ (where $\bar{V}_{\mathrm{mi}}$ stands for the partial molar volume of component i) and embedded in a dielectric continuum formed by the real mixture at the same composition. This approach leads to the definition of the relative excess Kirkwood correlation factor of liquid i, which is given by:

$$g_{\mathrm{K,rel,i}}^{\mathrm{E}} = \frac{\bar{V}_{\mathrm{mi}}(\varepsilon_{\mathrm{r}} - \varepsilon_{\mathrm{ri}}^{\infty})(2\varepsilon_{\mathrm{r}} + \varepsilon_{\mathrm{ri}}^{\infty})\varepsilon_{\mathrm{r}}^{\mathrm{id}}}{V_{\mathrm{mi}}(\varepsilon_{\mathrm{r}}^{\mathrm{id}} - \varepsilon_{\mathrm{r}}^{\infty})(2\varepsilon_{\mathrm{r}}^{\mathrm{id}} + \varepsilon_{\mathrm{ri}}^{\infty})\varepsilon_{\mathrm{r}}} - 1 \qquad (17)$$

Values of $g_{\mathrm{K,rel}}^{\mathrm{E}}$ and $g_{\mathrm{K,rel,i}}^{\mathrm{E}}$ are collected in Table 7 (Figures 7-10). From inspection of the results obtained some conclusions regarding systems with linear amines can be stated. (i) The $g_{\mathrm{K,rel}}^{\mathrm{E}}$ values are negative over the whole composition range. As in the ideal mixture neither correlations between like dipoles are destroyed nor are new correlations between unlike dipoles created, these results show that there is a destruction of the structure in the solution with regards to that of the ideal mixture. (ii) Interestingly, the $g_{\mathrm{K,rel}}^{\mathrm{E}}$ curves are skewed towards low $\phi_1$ values (Figures 7 and 8). This suggests that the amide structure is better destroyed at such concentrations. (iii) An interesting result is that the minima of the $g_{\mathrm{K,rel}}^{\mathrm{E}}$ curves is reached at lower volume fractions of the amide than in the $\varepsilon_{\mathrm{r}}^{\mathrm{E}}$ and $(\Pi_{\mathrm{m}}^{\mathrm{or}})^{\mathrm{E}}$ curves (Table 7). Thus, according to the Kirkwood-Fröhlich model, the destruction of dipole correlations is not the only responsible for the $\varepsilon_{\mathrm{r}}^{\mathrm{E}}$ minima, but other related effects, such as the number and strength of interactions created and disrupted upon mixing, are also important. (iv) The minimum values change in similar order to that encountered for $\varepsilon_{\mathrm{r}}^{\mathrm{E}}$. For example, $g_{\mathrm{K,rel}}^{\mathrm{E}}$ (DMA) = − 0.14 (BA) > − 0.18 (HxA) > − 0.21 (DPA) > − 0.24 (DBA). The BA mixture is the most structured, which can be ascribed to a higher relevance of the creation of interactions between unlike molecules. (v) Similarly, systems with DMF are also more structured than those with DMA. (vi) For a given mixture, the $g_{\mathrm{K,rel,i}}^{\mathrm{E}}$ values are practically independent of the component considered and are similar to $g_{\mathrm{K,rel}}^{\mathrm{E}}$ values (Table 7). This may mean than interaction between unlike molecules partially compensate the loss of structure of the components. (vi) The $\left|g_{\mathrm{K,rel,1}}^{\mathrm{E}}\right|$ values are larger for DMA than for DMF in systems with a given linear amine. That is, the loss of order in the liquid state is higher in the vicinity of a DMA molecule than around a DMF molecule.

Finally, we must remark the positive values of $g_{\mathrm{K,rel}}^{\mathrm{E}}$ and $g_{\mathrm{K,rel,i}}^{\mathrm{E}}$ for the DMF + aniline mixture. They show that the passage from an ideal to a real mixture leads here to an increment of the order in the liquid state, which, in addition, is higher in the neighbourhood of the aniline molecules. The DMF + HxA system behaves in the opposite way and there is a loss of order in the liquid state when passing from an ideal to a real mixture.

## 7. Conclusions

Measurements on $\varepsilon_{\mathrm{r}}$ have been reported for the systems: DMA + BA, + HxA, + DPA, or + DBA and for DMF + aniline at (293.15-303.15) K. The corresponding $\varepsilon_{\mathrm{r}}^{\mathrm{E}}$ values are large and negative for mixtures with linear amines and positive for the aniline solution. In the former case, this means that the main contributions to $\varepsilon_{\mathrm{r}}^{\mathrm{E}}$ come from the disruption of interactions between like molecules. In the latter case, the $\varepsilon_{\mathrm{r}}^{\mathrm{E}}$ sign is determined by the positive contribution from the DMF-aniline interactions. Inspection of $\varepsilon_{\mathrm{r}}^{\mathrm{E}}$ data shows that: (i) longer linear amines are better breakers of the amide-amide interactions; (ii) interactions between unlike molecules are more easily formed when shorter linear amines, or DMF, participate. Calculations on $(\Pi_{\mathrm{m}}^{\mathrm{or}})^{\mathrm{E}}$, $(\Pi_{\mathrm{m}}^{\mathrm{ind}})^{\mathrm{E}}$, $g_{\mathrm{K,rel}}^{\mathrm{E}}$ and $g_{\mathrm{K,rel,i}}^{\mathrm{E}}$, and the dependence of $K_{\mathrm{AB}}$ with the molecular structure are consistent with these findings.

## Acknowledgements

The authors gratefully acknowledge the financial support received from the Consejería de Educación y Cultura of Junta de Castilla y León, under Project BU034U16. F. Hevia and A. Cobos are grateful to Ministerio de Educación, Cultura y Deporte for the grants FPU14/04104 and FPU15/05456 respectively.

Table 1

Sample description.

| Chemical name | CAS Number | Source | Purification method | Purity[a] |
|---|---|---|---|---|
| *N*,*N*-dimethylacetamide (DMA) | 127-19-5 | Sigma-Aldrich | None | 0.9998 |
| *N*,*N*-dimethylformamide (DMF) | 68-12-2 | Sigma-Aldrich | None | 0.9995 |
| *N*-propylpropan-1-amine (DPA) | 142-84-7 | Aldrich | None | 0.996 |
| *N*-butylbutan-1-amine (DBA) | 111-92-2 | Aldrich | None | 0.9974 |
| Butan-1-amine (BA) | 109-73-9 | Sigma-Aldrich | None | 0.9996 |
| Hexan-1-amine (HxA) | 111-26-2 | Aldrich | None | 0.999 |
| Aniline | 62-53-3 | Sigma-Aldrich | None | 0.999 |

[a] In mole fraction. Provided by the supplier by gas chromatography.

Table 2

Relative permittivity, $\varepsilon_r^*$, of pure compounds at temperature *T*, pressure *p* = 0.1 MPa and frequency $\nu = 1$ MHz. [a]

| Compound[b] | *T*/K | $\varepsilon_r^*$ | | $V_m \frac{\partial \varepsilon_r^*}{\partial T}$ / cm³·mol⁻¹·K⁻¹ | $\mu$/D |
|---|---|---|---|---|---|
| | | Exp. | Lit. | | |
| DMA | 293.15 | 39.695 | | | |
| | 298.15 | 38.586 | 38.60 [59]; 43.00 [60], 37.78 [61] | – 20.47 [c] | 3.7 [2] |
| | 303.15 | 37.499 | 37.72 [62]; 38.67 [63] | | |
| DMF | 293.15 | 38.334 | 38.30 [64] | | |
| | 298.15 | 37.440 | 37.65 [65], 37.6 [50] | – 13.55 [d] | 3.7 [3] |
| | 303.15 | 36.580 | 36.55 [66] | | |
| DPA | 293.15 | 3.148 | 3.31 [67]; 3.068 [3] | | |
| | 298.15 | 3.093 | 3.24 [67] | – 1.52 [c] | 1.0 [15] |
| | 303.15 | 3.037 | 3.18 [3] | | |
| DBA | 293.15 | 2.938 | 2.978 [3]; 2.765 [68] | | |
| | 298.15 | 2.896 | | – 1.37 [c] | 1.1 [15] |
| | 303.15 | 2.858 | 2.697 [68] | | |
| BA | 293.15 | 4.729 | 4.71 [69]; 4.88 [3]; 4.91 [70]; 5.34 [71]; 4.70 [47] | | |
| | 298.15 | 4.636 | 4.62 [70]; 5.16 [71] | – 1.90 [c] | 1.3 [15] |
| | 303.15 | 4.547 | 4.57 [70]; 4.48 [71] | | |
| HxA | 293.15 | 3.955 | 3.94 [47] | | |
| | 298.15 | 3.893 | | – 1.73 [c] | 1.3 [2] |
| | 303.15 | 3.835 | 3.83 [47] | | |
| Aniline [e] | 293.15 | 7.117 | 6.48 [72]; 6.55 [73] | | |
| | 298.15 | 6.984 | 6.774 [74], 6.59 [75] | – 2.39 | 1.51 [3] |
| | 303.15 | 6.856 | 6.09 [72]; 6.0 [73]; 6.71 [3]; 6.88 [76]; 6.857 [77]; 6.055 [78] | | |

[a]The standard uncertainties are: $u(T) = 0.02$ K; $u(p) = 1$ kPa; $u(\nu) = 20$ Hz. The relative standard uncertainty is: $u_r(\varepsilon_r^*) = 0.003$; [b]for symbols, see Table 1; [c]molar volume, $V_m$, taken from ref. [24]; [d]$V_m$ taken from ref. [23]; [e] $V_m$ taken from Table S1.

Table 3

Volume fractions of amide, $\phi_1$, relative permittivities, $\varepsilon_\mathrm{r}$, and excess relative permittivities, $\varepsilon_\mathrm{r}^\mathrm{E}$, of DMA (1) + amine (2) and and DMF (1) + aniline (2)[a] mixtures as functions of the mole fraction of amide, $x_1$, at temperature *T*, pressure *p* = 0.1 MPa and frequency $\nu = 1$ MHz. [b]

| $x_1$ | $\phi_1$ | $\varepsilon_\mathrm{r}$ | $\varepsilon_\mathrm{r}^\mathrm{E}$ | $x_1$ | $\phi_1$ | $\varepsilon_\mathrm{r}$ | $\varepsilon_\mathrm{r}^\mathrm{E}$ |
|---|---|---|---|---|---|---|---|
| | | | DMA (1) + DPA (2) ; *T*/K = 293.15 | | | | |
| 0.0000 | 0.0000 | 3.148 | | 0.6398 | 0.5453 | 20.725 | – 2.352 |
| 0.0600 | 0.0413 | 4.165 | – 0.492 | 0.6985 | 0.6100 | 23.242 | – 2.200 |
| 0.1099 | 0.0769 | 5.107 | – 0.851 | 0.7479 | 0.6670 | 25.568 | – 1.957 |
| 0.1494 | 0.1060 | 5.897 | – 1.125 | 0.7948 | 0.7234 | 27.856 | – 1.730 |
| 0.2122 | 0.1539 | 7.264 | – 1.509 | 0.8464 | 0.7881 | 30.535 | – 1.416 |
| 0.3021 | 0.2261 | 9.460 | – 1.951 | 0.8928 | 0.8490 | 33.134 | – 1.042 |
| 0.4041 | 0.3140 | 12.353 | – 2.271 | 0.9494 | 0.9268 | 36.494 | – 0.526 |
| 0.4917 | 0.3951 | 15.141 | – 2.447 | 1.0000 | 1.0000 | 39.695 | |
| 0.5905 | 0.4933 | 18.754 | – 2.423 | | | | |
| | | | DMA (1) + DPA (2) ; *T*/K = 298.15 | | | | |
| 0.0000 | 0.0000 | 3.093 | | 0.6398 | 0.5450 | 20.156 | – 2.281 |
| 0.0600 | 0.0413 | 4.083 | – 0.476 | 0.6985 | 0.6097 | 22.613 | – 2.120 |
| 0.1099 | 0.0768 | 4.990 | – 0.829 | 0.7479 | 0.6667 | 24.852 | – 1.904 |
| 0.1494 | 0.1059 | 5.757 | – 1.095 | 0.7948 | 0.7231 | 27.082 | – 1.676 |
| 0.2122 | 0.1537 | 7.088 | – 1.460 | 0.8464 | 0.7879 | 29.694 | – 1.364 |
| 0.3021 | 0.2259 | 9.217 | – 1.894 | 0.8928 | 0.8488 | 32.205 | – 1.014 |
| 0.4041 | 0.3138 | 12.018 | – 2.213 | 0.9494 | 0.9267 | 35.464 | – 0.520 |
| 0.4917 | 0.3947 | 14.741 | – 2.361 | 1.0000 | 1.0000 | 38.586 | |
| 0.5905 | 0.4930 | 18.239 | – 2.352 | | | | |
| | | | DMA (1) + DPA (2) ; *T*/K = 303.15 | | | | |
| 0.0000 | 0.0000 | 3.037 | | 0.6398 | 0.5446 | 19.598 | – 2.207 |
| 0.0600 | 0.0412 | 3.999 | – 0.458 | 0.6985 | 0.6094 | 21.993 | – 2.045 |
| 0.1099 | 0.0768 | 4.873 | – 0.811 | 0.7479 | 0.6664 | 24.135 | – 1.867 |
| 0.1494 | 0.1058 | 5.619 | – 1.064 | 0.7948 | 0.7229 | 26.324 | – 1.626 |
| 0.2122 | 0.1535 | 6.912 | – 1.415 | 0.8464 | 0.7877 | 28.874 | – 1.309 |
| 0.3021 | 0.2257 | 8.980 | – 1.835 | 0.8928 | 0.8487 | 31.293 | – 0.992 |
| 0.4041 | 0.3135 | 11.690 | – 2.151 | 0.9494 | 0.9267 | 34.462 | – 0.511 |
| 0.4917 | 0.3944 | 14.338 | – 2.291 | 1.0000 | 1.0000 | 37.499 | |

TABLE 3 (continued)

| | | | | | | | |
|---|---|---|---|---|---|---|---|
| 0.5905 | 0.4926 | 17.732 | – 2.281 | | | | |
| | | | DMA (1) + DBA (2) ; *T*/K = 293.15 | | | | |
| 0.0000 | 0.0000 | 2.938 | | 0.6015 | 0.4510 | 16.820 | – 2.695 |
| 0.0896 | 0.0508 | 4.186 | – 0.619 | 0.6429 | 0.4949 | 18.472 | – 2.657 |
| 0.1507 | 0.0881 | 5.155 | – 1.021 | 0.7094 | 0.5706 | 21.390 | – 2.522 |
| 0.2190 | 0.1324 | 6.372 | – 1.433 | 0.7469 | 0.6163 | 23.221 | – 2.370 |
| 0.3109 | 0.1971 | 8.266 | – 1.917 | 0.7992 | 0.6842 | 25.981 | – 2.106 |
| 0.3974 | 0.2641 | 10.350 | – 2.296 | 0.8428 | 0.7448 | 28.506 | – 1.809 |
| 0.4353 | 0.2956 | 11.384 | – 2.419 | 0.8963 | 0.8247 | 31.920 | – 1.331 |
| 0.4940 | 0.3470 | 13.117 | – 2.576 | 0.9456 | 0.9044 | 35.389 | – 0.792 |
| 0.5505 | 0.4000 | 14.971 | – 2.670 | 1.0000 | 1.0000 | 39.695 | |
| | | | DMA (1) + DBA (2) ; *T*/K = 298.15 | | | | |
| 0.0000 | 0.0000 | 2.896 | | 0.6015 | 0.4509 | 16.376 | – 2.613 |
| 0.0896 | 0.0508 | 4.106 | – 0.603 | 0.6429 | 0.4948 | 17.967 | – 2.588 |
| 0.1507 | 0.0880 | 5.045 | – 0.992 | 0.7094 | 0.5704 | 20.804 | – 2.450 |
| 0.2190 | 0.1323 | 6.224 | – 1.394 | 0.7469 | 0.6161 | 22.565 | – 2.320 |
| 0.3109 | 0.1970 | 8.063 | – 1.864 | 0.7992 | 0.6840 | 25.246 | – 2.062 |
| 0.3974 | 0.2640 | 10.085 | – 2.233 | 0.8428 | 0.7446 | 27.715 | – 1.756 |
| 0.4353 | 0.2954 | 11.084 | – 2.355 | 0.8963 | 0.8246 | 31.031 | – 1.295 |
| 0.4940 | 0.3468 | 12.766 | – 2.507 | 0.9456 | 0.9043 | 34.398 | – 0.772 |
| 0.5505 | 0.3998 | 14.575 | – 2.590 | 1.0000 | 1.0000 | 38.586 | |
| | | | DMA (1) + DBA (2) ; *T*/K = 303.15 | | | | |
| 0.0000 | 0.0000 | 2.858 | | 0.6015 | 0.4507 | 15.942 | – 2.529 |
| 0.0896 | 0.0508 | 4.033 | – 0.585 | 0.6429 | 0.4946 | 17.504 | – 2.487 |
| 0.1507 | 0.0880 | 4.942 | – 0.964 | 0.7094 | 0.5703 | 20.253 | – 2.361 |
| 0.2190 | 0.1323 | 6.086 | – 1.355 | 0.7469 | 0.6160 | 21.966 | – 2.231 |
| 0.3109 | 0.1970 | 7.871 | – 1.811 | 0.7992 | 0.6839 | 24.579 | – 1.970 |
| 0.3974 | 0.2639 | 9.833 | – 2.167 | 0.8428 | 0.7446 | 26.984 | – 1.668 |
| 0.4353 | 0.2953 | 10.809 | – 2.278 | 0.8963 | 0.8245 | 30.209 | – 1.211 |
| 0.4940 | 0.3467 | 12.440 | – 2.428 | 0.9456 | 0.9043 | 33.491 | – 0.693 |
| 0.5505 | 0.3997 | 14.195 | – 2.509 | 1.0000 | 1.0000 | 37.499 | |
| | | | DMA (1) + BA (2) ; *T*/K = 293.15 | | | | |
| 0.0000 | 0.0000 | 4.729 | | 0.5989 | 0.5822 | 23.127 | – 1.935 |
| 0.0584 | 0.0547 | 6.126 | – 0.513 | 0.6947 | 0.6798 | 26.781 | – 1.689 |

TABLE 3 (continued)

| | | | | | | | |
|---|---|---|---|---|---|---|---|
| 0.1069 | 0.1005 | 7.365 | – 0.874 | 0.7906 | 0.7789 | 30.606 | – 1.325 |
| 0.1973 | 0.1866 | 9.841 | – 1.405 | 0.8404 | 0.8309 | 32.664 | – 1.083 |
| 0.3034 | 0.2890 | 13.000 | – 1.822 | 0.8970 | 0.8904 | 35.070 | – 0.755 |
| 0.4037 | 0.3872 | 16.238 | – 2.014 | 0.9491 | 0.9457 | 37.354 | – 0.403 |
| 0.4978 | 0.4805 | 19.477 | – 2.033 | 1.0000 | 1.0000 | 39.653 | |
| DMA (1) + BA (2) ; *T*/K = 298.15 | | | | | | | |
| 0.0000 | 0.0000 | 4.636 | | 0.5989 | 0.5818 | 22.511 | – 1.842 |
| 0.0584 | 0.0546 | 5.989 | – 0.497 | 0.6947 | 0.6795 | 26.066 | – 1.598 |
| 0.1069 | 0.1003 | 7.194 | – 0.841 | 0.7906 | 0.7786 | 29.779 | – 1.244 |
| 0.1973 | 0.1863 | 9.596 | – 1.354 | 0.8404 | 0.8307 | 31.779 | – 1.009 |
| 0.3034 | 0.2887 | 12.665 | – 1.755 | 0.8970 | 0.8903 | 34.105 | – 0.703 |
| 0.4037 | 0.3868 | 15.823 | – 1.922 | 0.9491 | 0.9456 | 36.315 | – 0.367 |
| 0.4978 | 0.4801 | 18.954 | – 1.953 | 1.0000 | 1.0000 | 38.526 | |
| DMA (1) + BA (2) ; *T*/K = 303.15 | | | | | | | |
| 0.0000 | 0.0000 | 4.547 | | 0.5989 | 0.5814 | 21.946 | – 1.769 |
| 0.0584 | 0.0545 | 5.863 | – 0.481 | 0.6947 | 0.6791 | 25.392 | – 1.544 |
| 0.1069 | 0.1002 | 7.034 | – 0.816 | 0.7906 | 0.7784 | 28.998 | – 1.211 |
| 0.1973 | 0.1861 | 9.380 | – 1.302 | 0.8404 | 0.8304 | 30.943 | – 0.981 |
| 0.3034 | 0.2883 | 12.352 | – 1.700 | 0.8970 | 0.8901 | 33.220 | – 0.672 |
| 0.4037 | 0.3864 | 15.431 | – 1.855 | 0.9491 | 0.9455 | 35.362 | – 0.356 |
| 0.4978 | 0.4797 | 18.478 | – 1.884 | 1.0000 | 1.0000 | 37.515 | |
| DMA (1) + HxA (2) ; *T*/K = 293.15 | | | | | | | |
| 0.0000 | 0.0000 | 3.955 | | 0.6944 | 0.6138 | 23.705 | – 2.183 |
| 0.1027 | 0.0741 | 5.810 | – 0.793 | 0.7581 | 0.6867 | 26.548 | – 1.945 |
| 0.1921 | 0.1426 | 7.690 | – 1.361 | 0.8028 | 0.7401 | 28.677 | – 1.724 |
| 0.3016 | 0.2320 | 10.349 | – 1.896 | 0.8528 | 0.8021 | 31.226 | – 1.390 |
| 0.3994 | 0.3175 | 13.099 | – 2.201 | 0.8986 | 0.8611 | 33.671 | – 1.054 |
| 0.5066 | 0.4180 | 16.508 | – 2.383 | 0.9529 | 0.9340 | 36.812 | – 0.518 |
| 0.6030 | 0.5151 | 20.011 | – 2.350 | 1.0000 | 1.0000 | 39.688 | |
| DMA (1) + HxA (2) ; *T*/K = 298.15 | | | | | | | |
| 0.0000 | 0.0000 | 3.893 | | 0.6944 | 0.6137 | 23.075 | – 2.118 |
| 0.1027 | 0.0741 | 5.694 | – 0.771 | 0.7581 | 0.6866 | 25.835 | – 1.888 |
| 0.1921 | 0.1425 | 7.522 | – 1.317 | 0.8028 | 0.7400 | 27.897 | – 1.679 |
| 0.3016 | 0.2319 | 10.096 | – 1.846 | 0.8528 | 0.8020 | 30.358 | – 1.370 |

TABLE 3 (continued)

| | | | | | | | |
|---|---|---|---|---|---|---|---|
| 0.3994 | 0.3173 | 12.772 | – 2.134 | 0.8986 | 0.8610 | 32.742 | – 1.034 |
| 0.5066 | 0.4178 | 16.083 | – 2.311 | 0.9529 | 0.9340 | 35.798 | – 0.511 |
| 0.6030 | 0.5150 | 19.493 | – 2.274 | 1.0000 | 1.0000 | 38.600 | |
| DMA (1) + HxA (2) ; *T*/K = 303.15 | | | | | | | |
| 0.0000 | 0.0000 | 3.835 | | 0.6944 | 0.6135 | 22.498 | – 2.022 |
| 0.1027 | 0.0740 | 5.588 | – 0.742 | 0.7581 | 0.6864 | 25.163 | – 1.815 |
| 0.1921 | 0.1424 | 7.362 | – 1.274 | 0.8028 | 0.7398 | 27.182 | – 1.597 |
| 0.3016 | 0.2317 | 9.866 | – 1.781 | 0.8528 | 0.8019 | 29.567 | – 1.306 |
| 0.3994 | 0.3172 | 12.465 | – 2.065 | 0.8986 | 0.8609 | 31.893 | – 0.969 |
| 0.5066 | 0.4177 | 15.690 | – 2.229 | 0.9529 | 0.9339 | 34.820 | – 0.503 |
| 0.6030 | 0.5148 | 19.005 | – 2.188 | 1.0000 | 1.0000 | 37.552 | |
| DMF (1) + aniline (2) ; *T*/K = 293.15 | | | | | | | |
| 0.0000 | 0.0000 | 7.117 | | 0.5999 | 0.5589 | 26.184 | 1.620 |
| 0.0533 | 0.0454 | 8.910 | 0.376 | 0.6989 | 0.6624 | 28.955 | 1.160 |
| 0.1046 | 0.0899 | 10.653 | 0.730 | 0.7931 | 0.7641 | 31.604 | 0.634 |
| 0.1562 | 0.1353 | 12.414 | 1.073 | 0.8431 | 0.8195 | 33.120 | 0.421 |
| 0.2033 | 0.1774 | 14.004 | 1.349 | 0.8982 | 0.8818 | 34.833 | 0.189 |
| 0.3015 | 0.2673 | 17.269 | 1.808 | 0.9459 | 0.9366 | 36.439 | 0.084 |
| 0.4071 | 0.3672 | 20.625 | 2.045 | 1.0000 | 1.0000 | 38.334 | |
| 0.5013 | 0.4593 | 23.428 | 1.973 | | | | |
| DMF (1) + aniline (2) ; *T*/K = 298.15 | | | | | | | |
| 0.0000 | 0.0000 | 6.984 | | 0.5999 | 0.5592 | 25.601 | 1.586 |
| 0.0533 | 0.0455 | 8.729 | 0.359 | 0.6989 | 0.6626 | 28.310 | 1.146 |
| 0.1046 | 0.0899 | 10.426 | 0.704 | 0.7931 | 0.7643 | 30.920 | 0.658 |
| 0.1562 | 0.1354 | 12.144 | 1.036 | 0.8431 | 0.8197 | 32.402 | 0.453 |
| 0.2033 | 0.1775 | 13.689 | 1.299 | 0.8982 | 0.8819 | 34.071 | 0.228 |
| 0.3015 | 0.2675 | 16.874 | 1.743 | 0.9459 | 0.9367 | 35.620 | 0.108 |
| 0.4071 | 0.3674 | 20.146 | 1.972 | 1.0000 | 1.0000 | 37.440 | |
| 0.5013 | 0.4596 | 22.895 | 1.913 | | | | |
| DMF (1) + aniline (2) ; *T*/K = 303.15 | | | | | | | |
| 0.0000 | 0.0000 | 6.856 | | 0.5999 | 0.5593 | 25.026 | 1.545 |
| 0.0533 | 0.0455 | 8.557 | 0.349 | 0.6989 | 0.6627 | 27.687 | 1.133 |
| 0.1046 | 0.0900 | 10.200 | 0.669 | 0.7931 | 0.7644 | 30.246 | 0.669 |
| 0.1562 | 0.1355 | 11.876 | 0.992 | 0.8431 | 0.8198 | 31.690 | 0.466 |

TABLE 3 (continued)

| | | | | | | | |
|---|---|---|---|---|---|---|---|
| 0.2033 | 0.1777 | 13.383 | 1.245 | 0.8982 | 0.8819 | 33.315 | 0.245 |
| 0.3015 | 0.2676 | 16.485 | 1.675 | 0.9459 | 0.9367 | 34.821 | 0.123 |
| 0.4071 | 0.3676 | 19.678 | 1.895 | 1.0000 | 1.0000 | 36.580 | |
| 0.5013 | 0.4597 | 22.371 | 1.851 | | | | |

[a]For symbols, see Table 1; [b]The standard uncertainties are: $u(T) = 0.02$ K; $u(p) = 1$ kPa; $u(\nu) = 20$ Hz; $u(x_1) = 0.0010$; $u(\phi_1) = 0.004$. The relative standard uncertainty is: $u_r(\varepsilon_r) = 0.003$; and the relative combined expanded uncertainty (0.95 level of confidence) is $U_{rc}(\varepsilon_r^E) = 0.03$.

Table 4

Coefficients $A_i$ and standard deviations, $\sigma\left(\varepsilon_r^E\right)$ (equation (4)), for the representation of $\varepsilon_r^E$ at temperature $T$ and pressure $p$ = 0.1 MPa for DMA (1) + amine (2) and DMF (1) + aniline (2) systems by equation (3).

| System[a] | $T$/K | $A_0$ | $A_1$ | $A_2$ | $A_3$ | $A_4$ | $\sigma\left(\varepsilon_r^E\right)$ |
|---|---|---|---|---|---|---|---|
| DMA + DPA | 293.15 | – 9.79 | – 1.40 | | | | 0.012 |
| | 298.15 | – 9.50 | – 1.34 | | | | 0.007 |
| | 303.15 | – 9.21 | – 1.30 | | | | 0.007 |
| DMA + DBA | 293.15 | – 10.35 | – 4.09 | – 1.03 | | | 0.011 |
| | 298.15 | – 10.06 | – 4.00 | – 1.06 | | | 0.009 |
| | 303.15 | – 9.76 | – 3.70 | – 0.69 | | | 0.006 |
| DMA + BA | 293.15 | – 8.16 | 0.72 | – 0.81 | | | 0.008 |
| | 298.15 | – 7.81 | 0.86 | – 0.64 | | | 0.006 |
| | 303.15 | – 7.53 | 0.83 | – 0.67 | | | 0.008 |
| DMA + HxA | 293.15 | – 9.49 | – 1.70 | – 0.9 | | | 0.010 |
| | 298.15 | – 9.19 | – 1.68 | – 1.0 | | | 0.012 |
| | 303.15 | – 8.86 | – 1.52 | – 0.9 | | | 0.011 |
| DMF + aniline | 293.15 | 7.87 | – 4.1 | – 5.6 | 1.0 | 1.7 | 0.012 |
| | 298.15 | 7.63 | – 3.83 | – 5.2 | 1.2 | 1.7 | 0.010 |
| | 303.15 | 7.38 | – 3.48 | – 4.8 | 1.2 | 1.7 | 0.009 |

[a]For symbols, see Table 1

Table 5

ERAS parameters[a] for amine (1) + amide (2) mixtures at 298.15 K.

| System[b] | $K_{AB}$ | $\Delta h^{*}_{AB}$ / kJ·mol$^{-1}$ | $\Delta v^{*}_{AB}$ / cm$^{3}$·mol$^{-1}$ | $X_{AB}$ / J·cm$^{-3}$ |
|---|---|---|---|---|
| BA + DMF | 1.2 | −22.0 | −2.5 | 10 |
| HxA + DMF | 0.65 | −22.0 | −2.5 | 10 |
| DPA + DMF | 0.6 | −22.0 | −3.1 | 10 |
| DBA + DMF | 0.20 | −22.0 | −3.1 | 17.45 |
| BA + DMA | 0.75 | −22.0 | −2.5 | 10 |
| HxA + DMA | 0.50 | −22.0 | −2.5 | 10 |
| DPA + DMA | 0.25 | −22.0 | −3.9 | 10 |
| DBA + DMA | 0.12 | −22.0 | −3.9 | 17.45 |
| Aniline + DMF | 70 | −22.0 | −11.1 | 4 |
| Aniline + DMA | 2.20 | −22.0 | −20 | 3.2 |
| HxA + NMA[c] | 20 | −25. | −3.2 | 5 |

[a] $K_{AB}$, association constant of component A with component B; $\Delta h^{*}_{AB}$, association enthalpy of component A with component B; $\Delta v^{*}_{AB}$, association volume of component A with component B; $X_{AB}$, physical parameter; [b]for symbols, see Table 1; [c]$T$ = 363.15 K.

Table 6

Excess molar volumes, $V_m^E$, at 298.15, equimolar composition and 0.1 MPa for amine (1) + *N*,*N*-dialkylamide (2) mixtures. Comparison of experimental results (Exp.) with ERAS calculations; the physical ($V_{m,phys}^E$) and chemical ($V_{m,chem}^E$) contributions are also listed.

| System[a] | $V_m^E$ | | | | Ref. |
|---|---|---|---|---|---|
| | Exp. | ERAS[b] | $V_{m,phys}^E$ | $V_{m,chem}^E$ | |
| BA + DMA | – 0.192 | – 0.208 | – 0.177 | – 0.031 | [24] |
| HxA + DMA | 0.006 | 0.006 | 0.043 | – 0.036 | [24] |
| DPA + DMA | – 0.227 | – 0.232 | – 0.242 | 0.01 | [24] |
| DBA + DMA | 0.056 | 0.051 | 0.116 | – 0.065 | [24] |
| BA + DMF | – 0.263 | – 0.270 | – 0.132 | – 0.139 | [23] |
| HxA + DMA | – 0.021 | – 0.024 | 0.075 | – 0.099 | [23] |
| DPA + DMF | – 0.289 | – 0.291 | – 0.181 | – 0.110 | [23] |
| DBA + DMF | 0.018 | 0.015 | 0.126 | – 0.111 | [23] |
| Aniline + DMA | – 0.609[c] | – 0.634 | 0.264 | – 0.898 | [45] |
| Aniline + DMF | – 0.662 | – 0.662 | 0.241 | – 0.903 | [27] |
| | – 0.693 | | | | [79] |

[a]For symbols, see Table 1; [b]results using ERAS parameters from Table 5; [c]$T$ = 303.15 K; $X_{AB}$ = 5 J·cm$^{-3}$.

Table 7

Excess functions, permittivity, $\varepsilon_r^E$, orientational polarizability, $(\Pi_m^{or})^E$, and relative Kirkwood's correlation factors ($g_{K,rel}^E$, $g_{K,rel,i}^E$, i = 1,2) for *N*,*N*-dialkylamide (1) + amine (2) mixtures at $\phi_1$ = 0.5 and 298.15 K. The minimum and maximum values of $\varepsilon_r^E$, $g_{K,rel}^E$, $g_{K,rel,i}^E$ ( i = 1,2) and the corresponding compositions are also listed.

| System[a] | $\varepsilon_r^E$ | $(\Pi_m^{or})^E$ / $cm^3 \cdot mol^{-1}$ | $g_{K,rel}^E$ | $g_{K,rel,1}^E$ | $g_{K,rel,2}^E$ |
|---|---|---|---|---|---|
| | | $\phi_1 = 0.5$ | | | |
| DMA + BA | − 1.943 | − 42.6 | − 0.10 | − 0.10 | − 0.10 |
| DMA + HxA | − 2.305 | − 56.4 | − 0.12 | − 0.12 | − 0.12 |
| DMA + DPA | − 2.348 | − 59.2 | − 0.12 | − 0.12 | − 0.12 |
| DMA + DBA | − 2.586 | − 69.5 | − 0.13 | − 0.13 | − 0.13 |
| DMF + BA | − 0.864 | − 18.0 | − 0.05 | − 0.05 | − 0.05 |
| DMF + HxA | − 1.262 | − 27.8 | − 0.07 | − 0.07 | − 0.06 |
| DMF + DPA | − 1.372 | − 31.6 | − 0.08 | − 0.08 | − 0.08 |
| DMF + DBA | − 1.733 | − 41.4 | − 0.09 | − 0.09 | − 0.09 |
| DMF + aniline | 1.806 | 30.5 | 0.10 | 0.08 | 0.08 |

Minimum or maximum values

| | $\phi_1$ | $\varepsilon_r^E$ | $\phi_1$ | $g_{K,rel}^E$ | $\phi_1$ | $g_{K,relA}^E$ | $\phi_1$ | $g_{K,relB}^E$ |
|---|---|---|---|---|---|---|---|---|
| DMA + BA | 0.45 | − 1.98 | 0.19 | − 0.14 | 0.17 | − 0.15 | 0.18 | − 0.14 |
| DMA + HxA | 0.46 | − 2.32 | 0.20 | − 0.18 | 0.15 | − 0.19 | 0.15 | − 0.18 |
| DMA + DPA | 0.43 | − 2.39 | 0.17 | − 0.21 | 0.11 | − 0.22 | 0.12 | − 0.21 |
| DMA + DBA | 0.45 | − 2.62 | 0.18 | − 0.24 | 0.10 | − 0.24 | 0.11 | − 0.24 |
| DMF + BA | 0.35 | − 0.97 | 0.20 | − 0.09 | 0.15 | − 0.09 | 0.15 | − 0.09 |
| DMF + HxA | 0.36 | − 1.38 | 0.22 | − 0.13 | 0.14 | − 0.13 | 0.14 | − 0.13 |
| DMF + DPA | 0.37 | − 1.51 | 0.22 | − 0.15 | 0.12 | − 0.16 | 0.14 | − 0.15 |
| DMF + DBA | 0.38 | − 1.86 | 0.26 | − 0.18 | 0.14 | − 0.18 | 0.14 | − 0.18 |
| DMF + aniline | 0.39 | 1.976 | 0.25 | 0.12 | 0.28 | 0.11 | 0.25 | 0.13 |

[a]For symbols, see Table 1

Table 8

Values of $V_{\mathrm{m}} \dfrac{\partial \varepsilon_{\mathrm{r}}^{*}}{\partial T}$ /cm$^3$·mol$^{-1}$·K$^{-1}$ at 298.15 K for *N*,*N*-dialkylamide (1) + amine (2) mixtures at 298.15 K and $\phi_1$ =0.5

| Amide | BA | HxA | DPA | DBA | Aniline |
|---|---|---|---|---|---|
| DMA | –9.86 | –10.9 | –11.1 | –11.6 | |
| DMF | –8.27 | –8.80 | –8.90 | –8.81 | –9.31 |

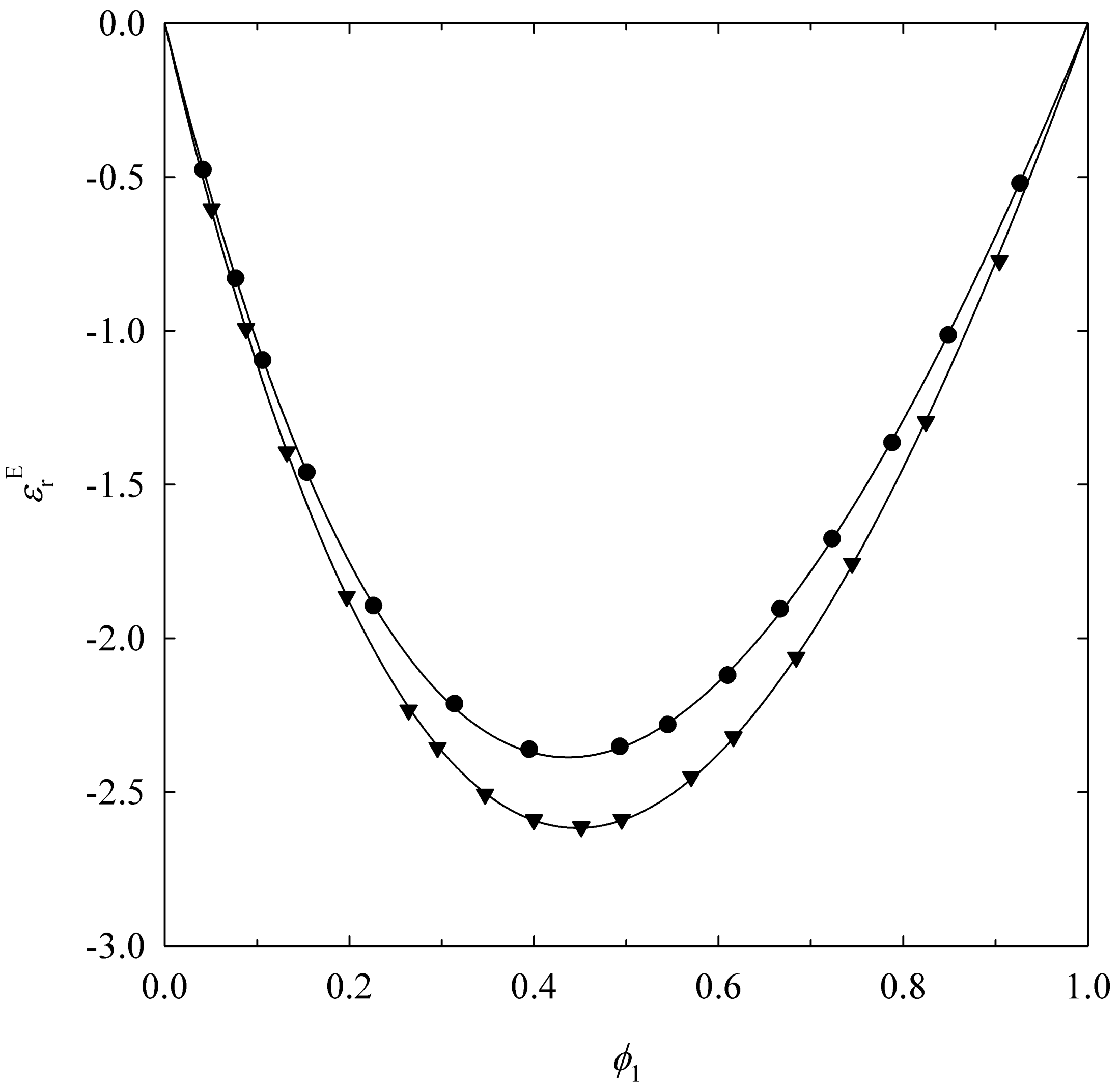

Figure 1

Excess relative permittivities, $\varepsilon_r^E$, for DMA (1) + DPA (2), or + DBA (2) systems at 0.1 MPa, 298.15 K and 1 MHz. Full symbols, experimental values (this work): (●), DPA; (▼), DBA. Solid lines, calculations with equation (3) using the coefficients from Table 4.

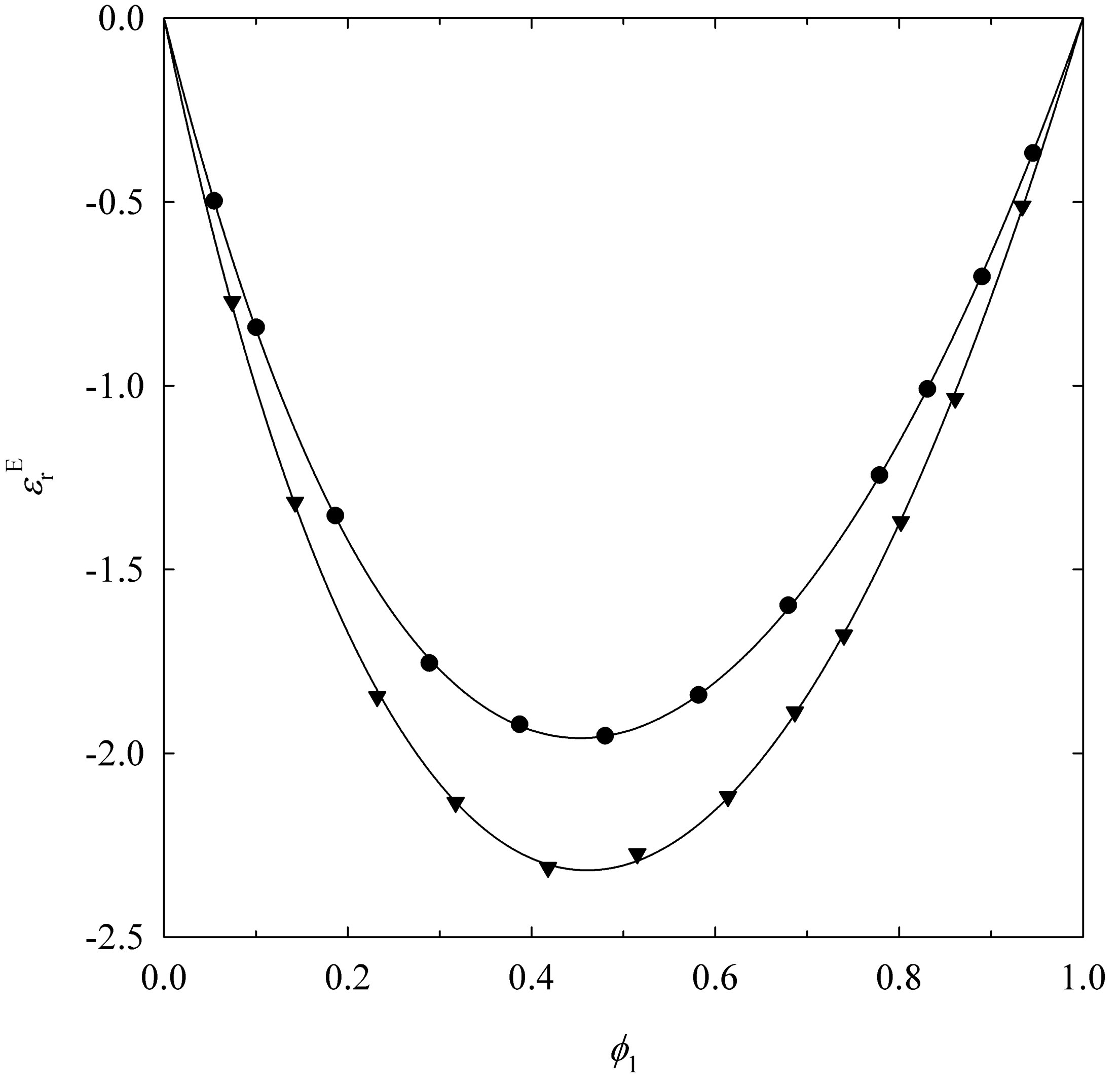

Figure 2

Excess relative permittivities, $\varepsilon_r^E$, for DMA (1) + BA (2), or + HxA (2) systems at 0.1 MPa, 298.15 K and 1 MHz. Full symbols, experimental values (this work): (●), BA; (▼), HxA. Solid lines, calculations with equation (3) using the coefficients from Table 4.

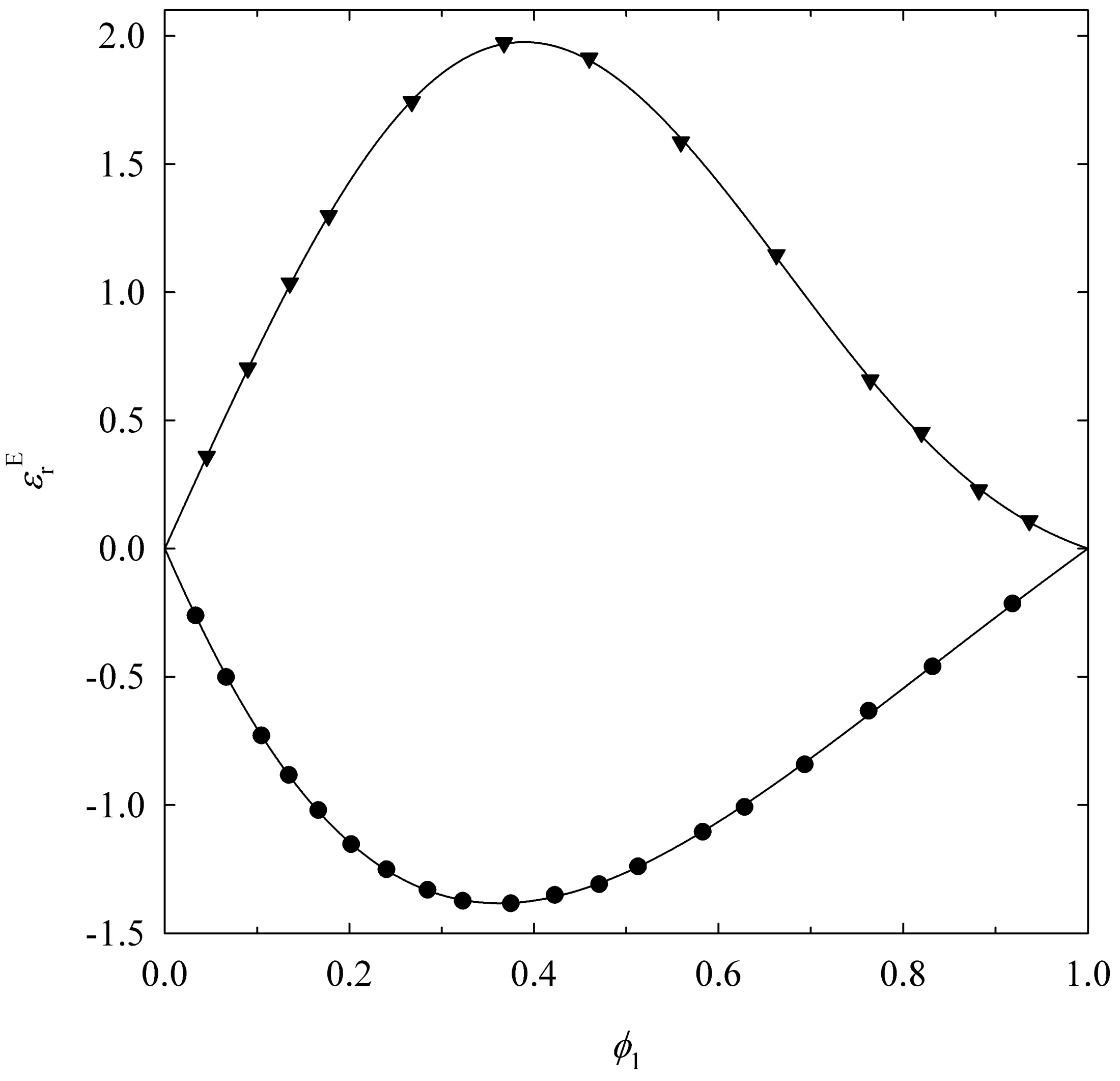

Figure 3

Excess relative permittivities, $\varepsilon_r^E$, for DMF (1) + amine (2) systems at 0.1 MPa, 298.15 K and 1 MHz. Full symbols, experimental values: (▼), aniline (this work); (●), HxA [25]. Solid lines, calculations with equation (3) using the coefficients from Table 4 or from the literature [25].

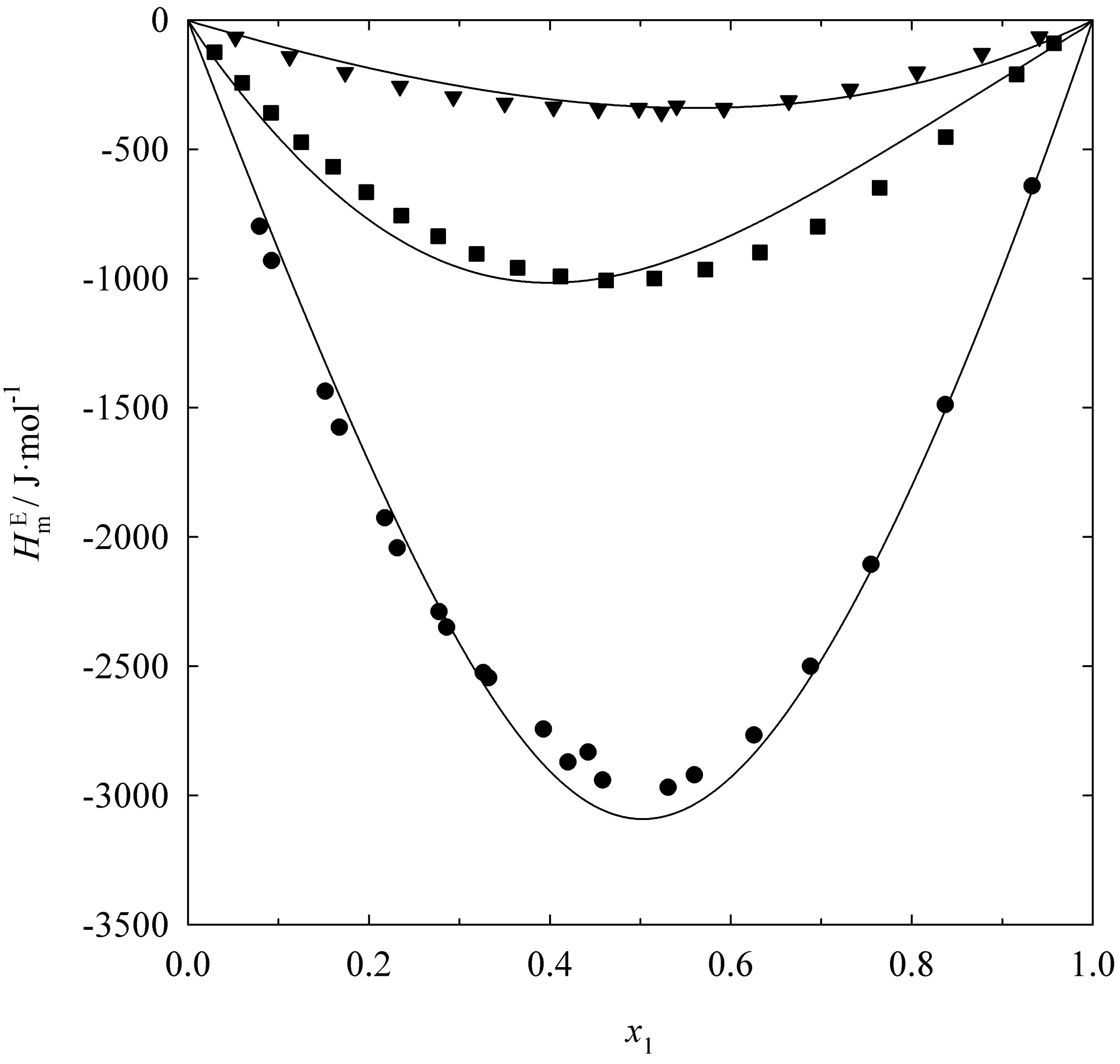

Figure 4

Excess molar enthalpies, $H_m^E$ for amine (1) + amide (2) mixtures at 0.1 MPa. Points experimental results: (●), aniline (1) + DMF (2) ($T$ = 298.15 K) [18]; (■), HxA (1) + NMA (2) ($T$ =363.15 K) [20]; (▲), aniline (1) + DMA (2) ($T$ = 298.15 K) [19]. Solid lines, ERAS calculations with parameters from Table 5.

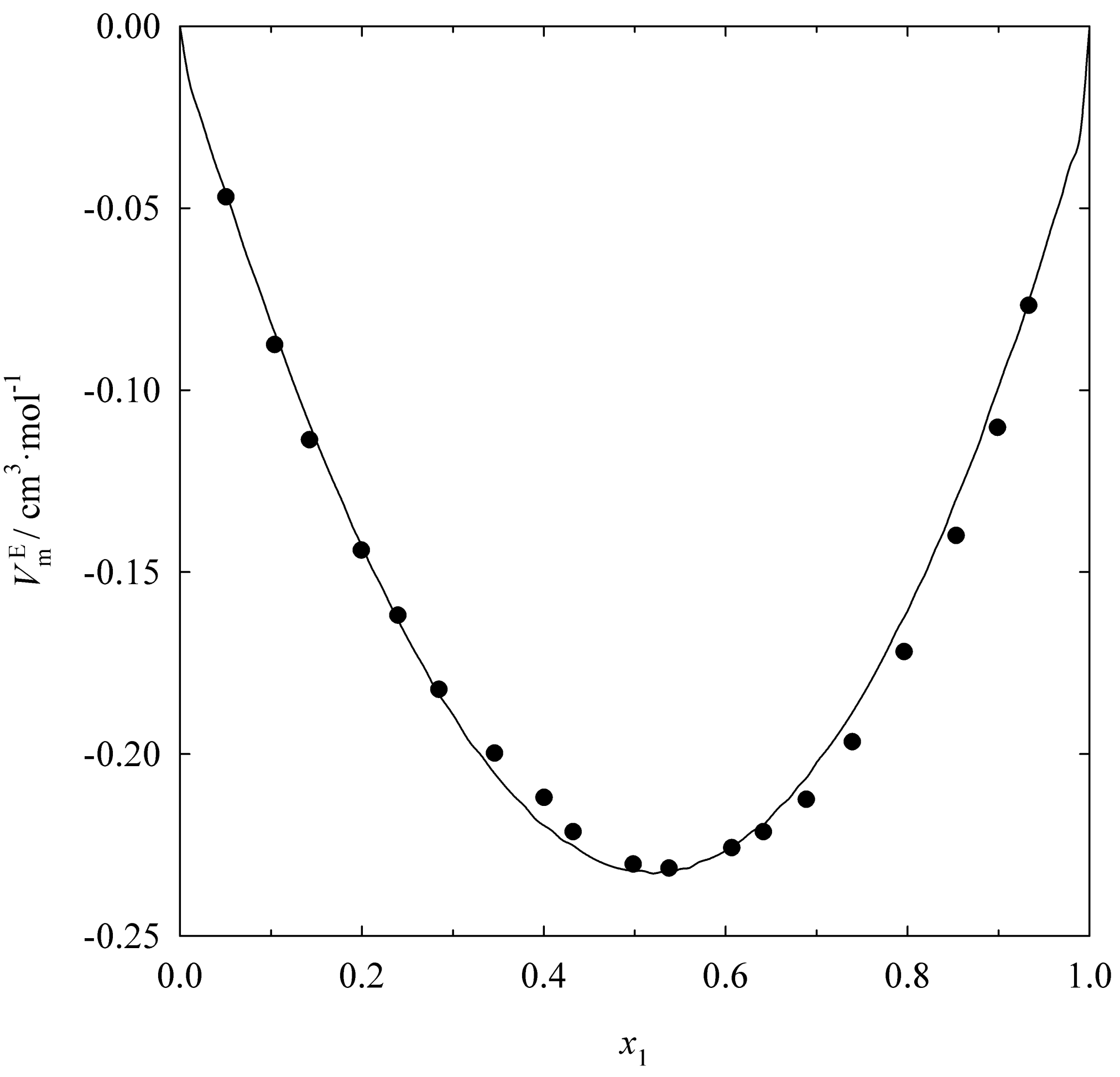

Figure 5

Excess molar volume, $V_m^E$ for DPA (1) + DMA (2) mixture at 298.15 K and 0.1 MPa. Points experimental results [24]. Solid line, ERAS calculations with parameters from Table 5.

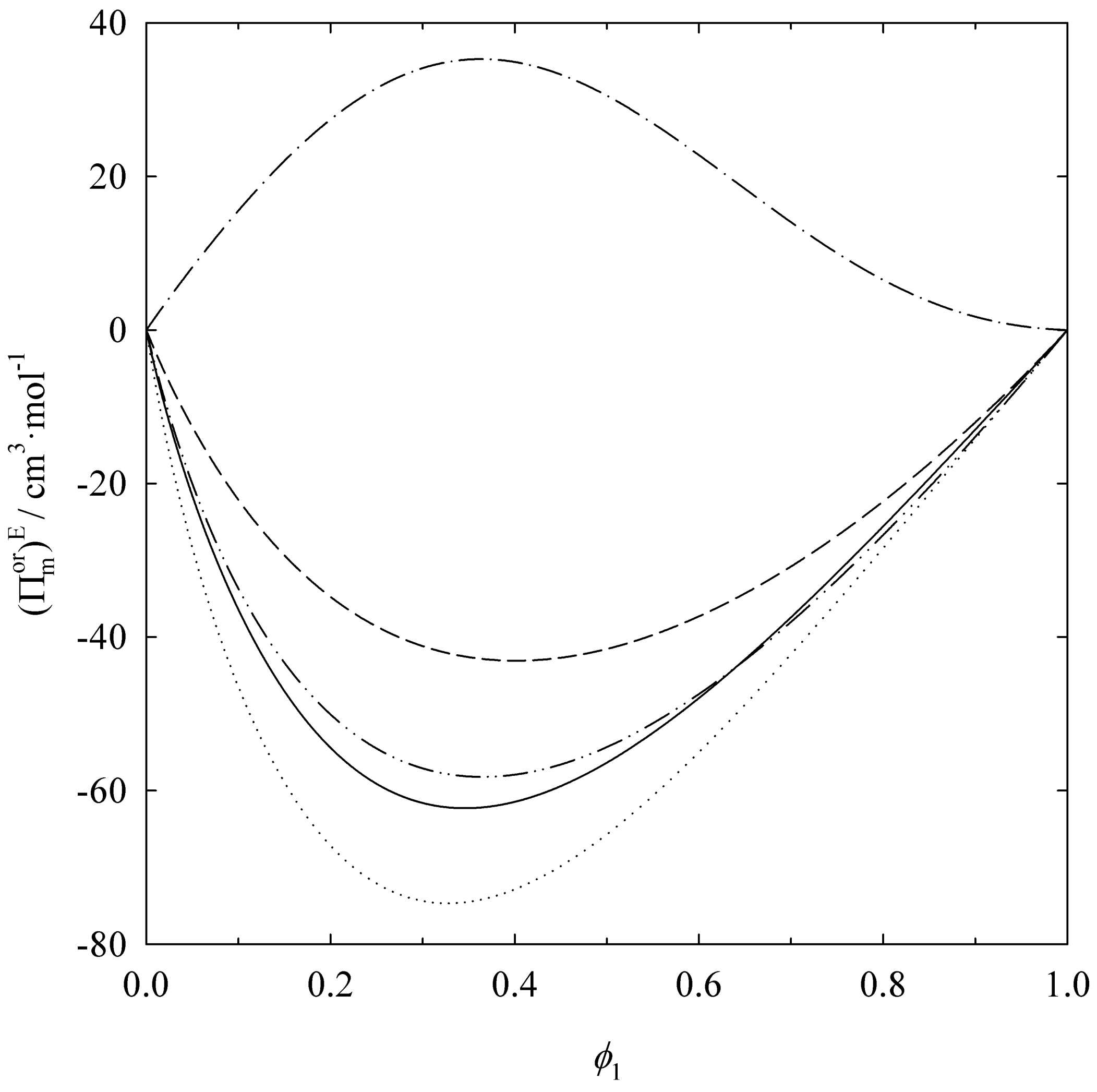

Figure 6

Excess molar orientational polarizability, $(\Pi_m^{or})^E$, for DMA (1) + linear amine (2), or DMF (1) + aniline (2) systems at 0.1 MPa and 298.15 K. (——), DMA + DPA; (···), DMA + DBA; (– – –), DMA + BA; (··–··), DMA + HxA; (·–·), DMF + aniline.

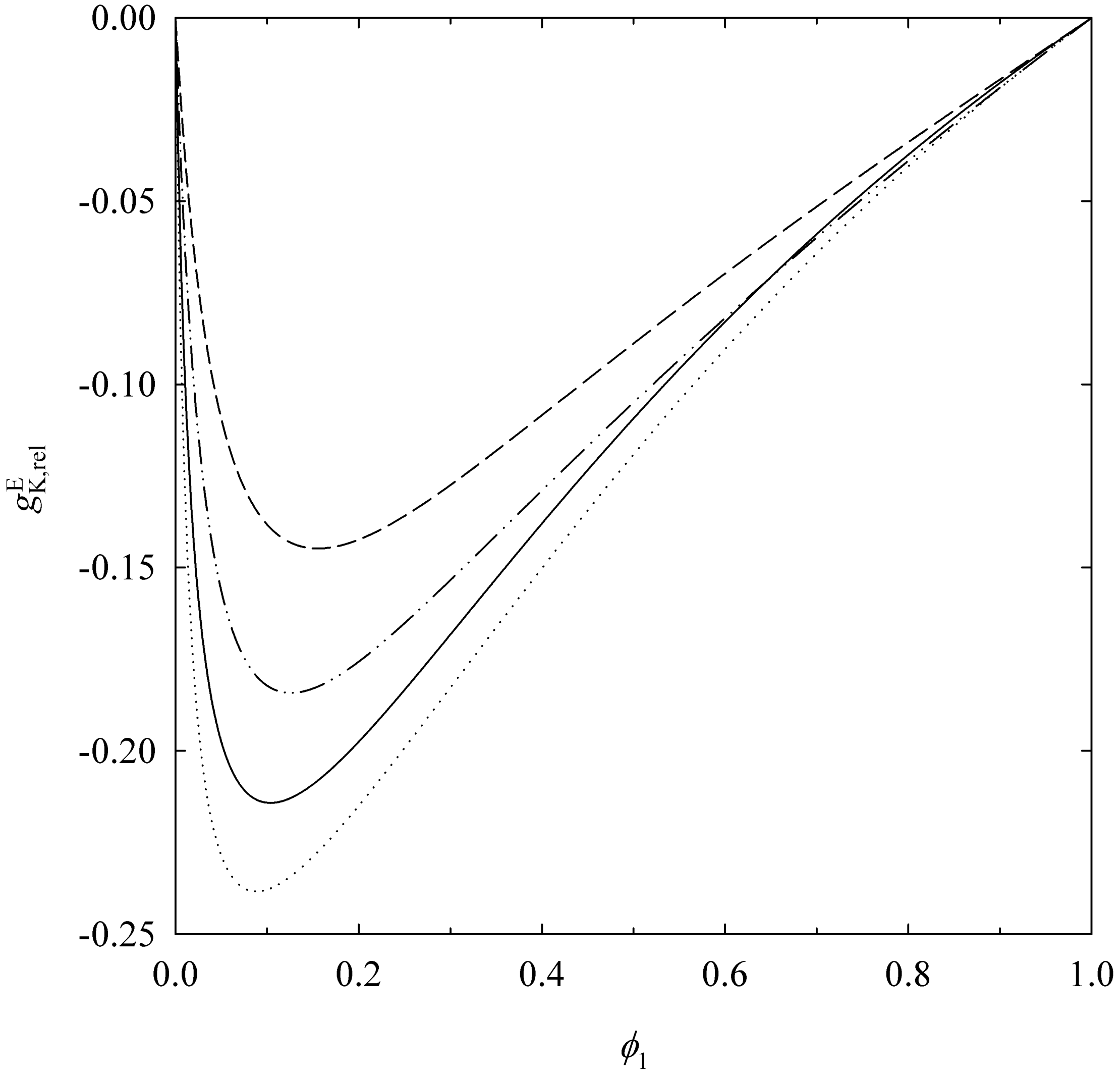

Figure 7

Excess relative Kirkwood correlation factors, $g^{E}_{K,rel}$, of DMA (1) + amine (2) systems at 0.1 MPa and 298.15 K. (——), DMA + DPA; (···), DMA + DBA; (– – –), DMA + BA; (··–··), DMA + HxA.

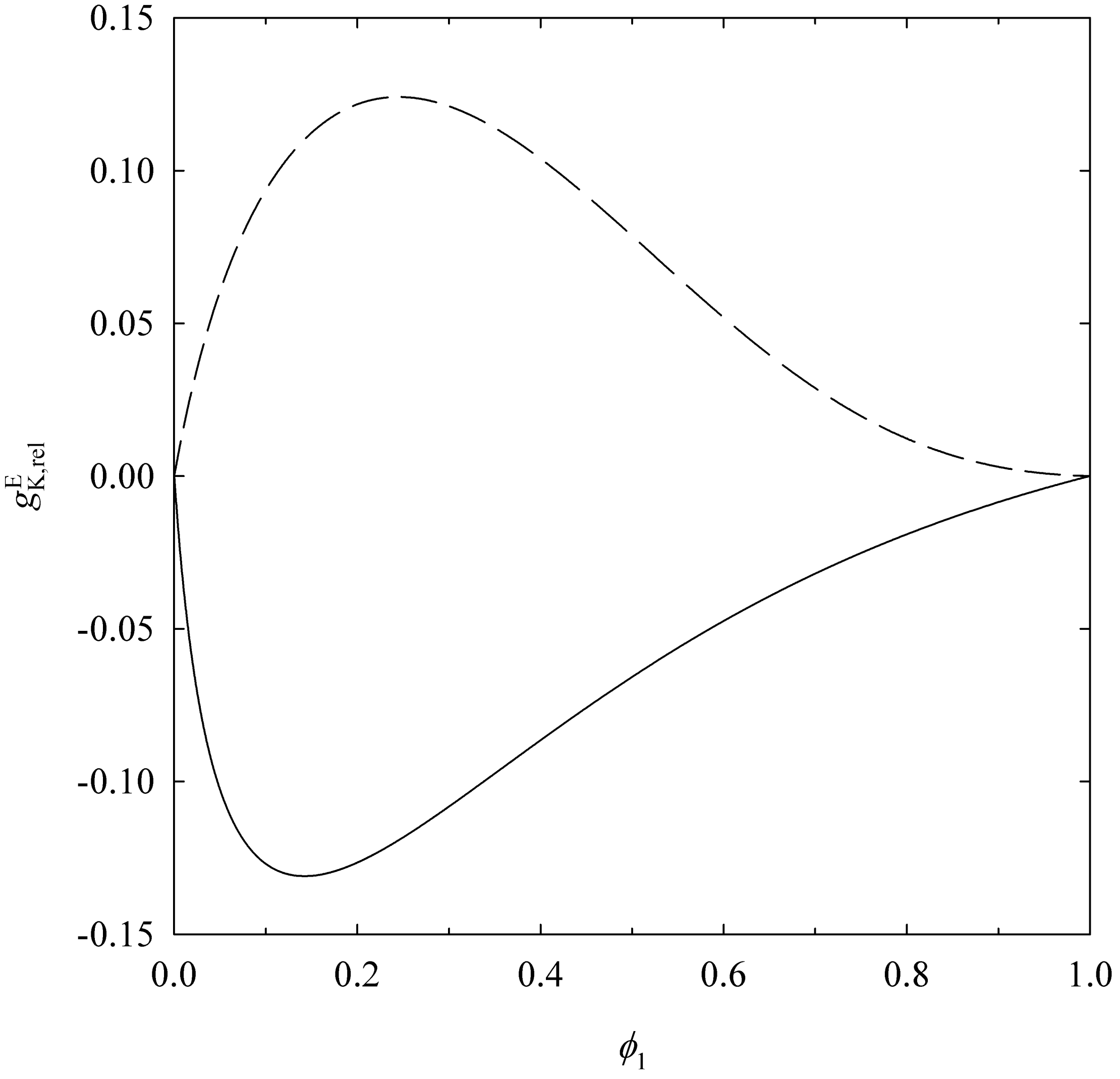

Figure 8

Excess relative Kirkwood correlation factors, $g^{E}_{K,rel}$, of DMF (1) + amine (2) systems at 0.1 MPa and 298.15 K. (——), HxA [25]; (– – –), aniline (this work).

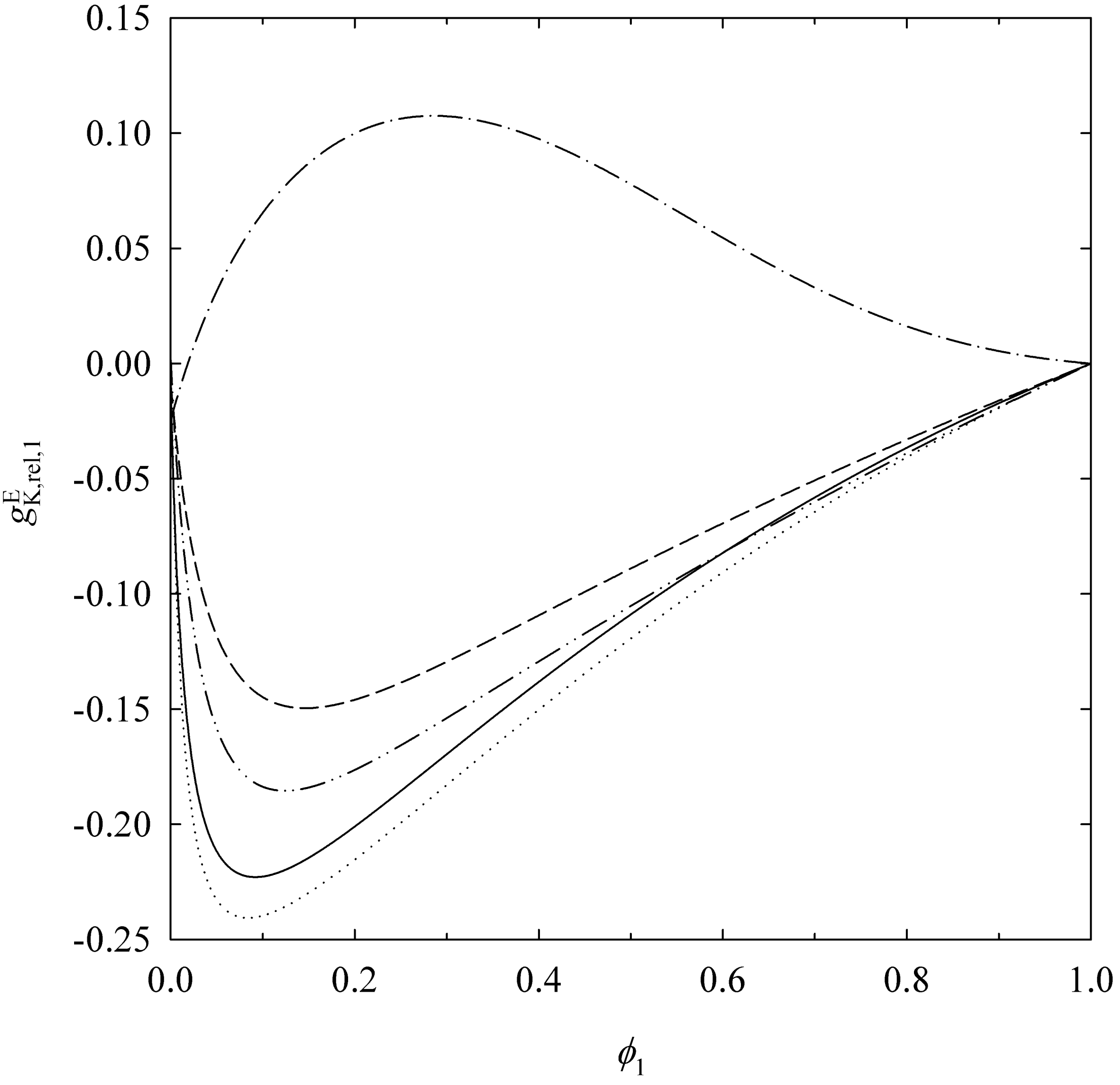

Figure 9

Excess relative Kirkwood correlation factors of liquid 1, $g^{E}_{K,rel,1}$, for DMA (1) + linear amine (2), or DMF (1) + aniline (2) systems at 0.1 MPa and 298.15 K. (––), DMA + DPA; (···), DMA + DBA; (– – –), DMA + BA; (··–··), DMA + HxA; (·–·), DMF + aniline.

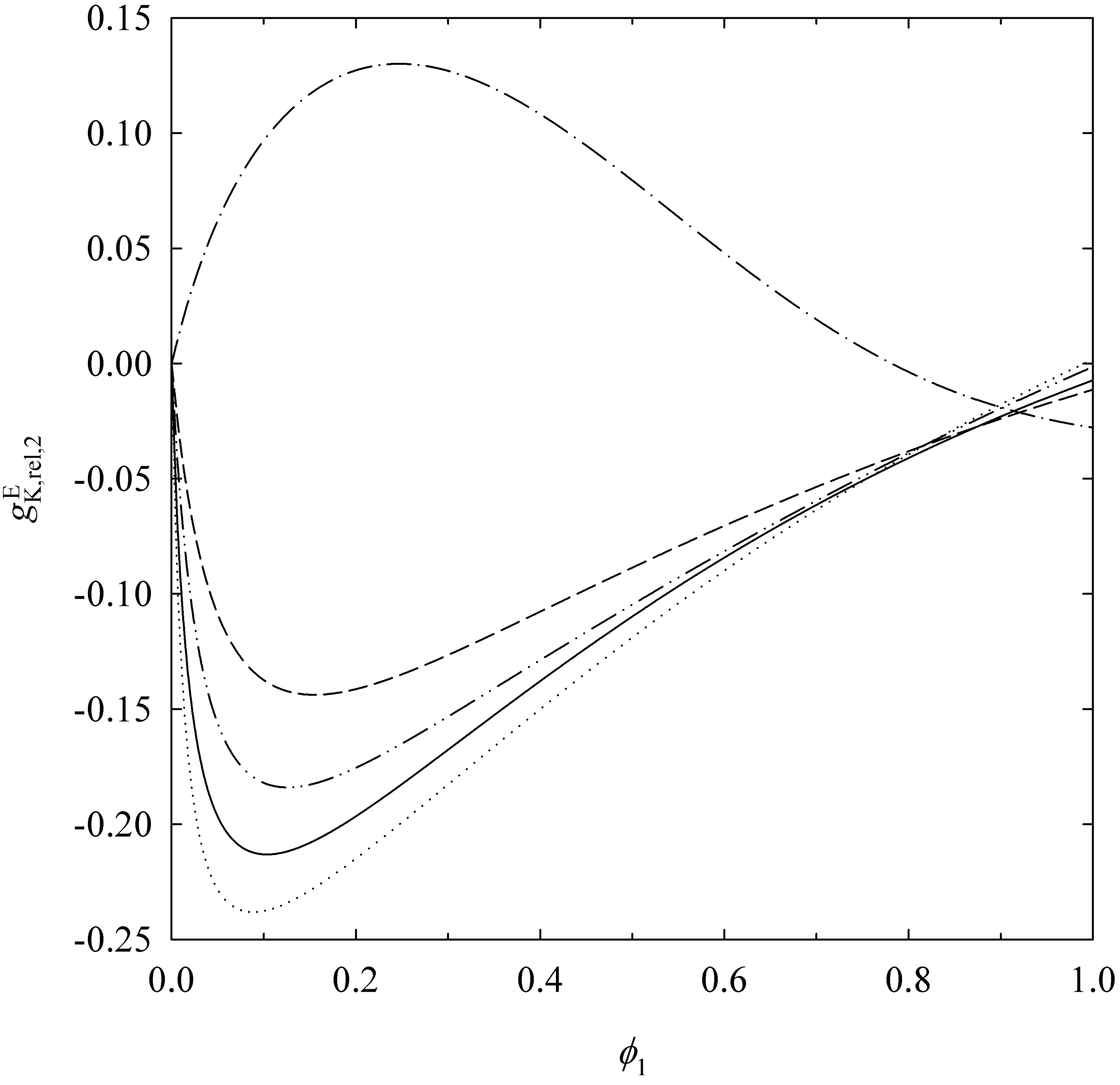

Figure 10

Excess relative Kirkwood correlation factors of liquid 2, $g^{E}_{K,rel,2}$, for DMA (1) + linear amine (2), or DMF (1) + aniline (2) systems at 0.1 MPa and 298.15 K. (––), DMA + DPA; (···), DMA + DBA; (– – –), DMA + BA; (··–··), DMA + HxA; (·–·), DMF + aniline.

**SUPPLEMENTARY MATERIAL**

# Thermodynamics of amide + amine mixtures. 4. Relative permittivities of *N*,*N*-dimethylacetamide + *N*-propylpropan-1-amine, + *N*-butylbutan-1-amine, + butan-1-amine, or + hexan-1-amine systems and of *N*,*N*-dimethylformamide + aniline mixture at several temperatures. Characterization of amine + amide systems using ERAS

Fernando Hevia, Juan Antonio González*, Ana Cobos, Isaías García de la Fuente, Luis Felipe Sanz

G.E.T.E.F., Departamento de Física Aplicada, Facultad de Ciencias, Universidad de Valladolid, Paseo de Belén, 7, 47011 Valladolid, Spain.

*e-mail: jagl@termo.uva.es; Fax: +34-983-423136; Tel: +34-983-423757

Table S1

Density, $\rho$, of aniline at temperature $T$ and 0.1 MPa[a]

| $T$/K | $\rho$ /g·cm$^{-3}$ | |
|---|---|---|
| | Experimental | Literature |
| 293.15 | 1.02179 | 1.0217 [s1] |
| | | 1.0173 [s2] |
| | | 1.02104 [s3] |
| 298.15 | 1.01752 | 1.0174 [s1] |
| | | 1.01744 [s4] |
| | | 1.0175 [s5] |
| | | 1.0176 [s6] |
| 303.15 | 1.01319 | 1.0130 [s1] |
| | | 1.01309 [s2] |
| | | 1.01318 [s7] |

[a] Density values measured by means of a vibrating-tube densimeter and sound analyser Anton Paar model DSA-5000. For details regarding the experimental method, see [s8,s9].

Standard uncertainties, $u$, are: $u(\rho) = 0.0012\rho$; $u(T) = 0.01$ K and for pressure, $u(p) = 1$ kPa.

Table S2

ERAS parameters[a] for pure compounds at 298.15 K.

| Compound | $V_{mi}$ / $cm^3 \cdot mol^{-1}$ | $\alpha_{Pi}$ / $10^3 \cdot K^{-1}$ | $\kappa_{Ti}$ / $TPa^{-1}$ | $K_i$ | $\Delta h_i^*$ / $kJ \cdot mol^{-1}$ | $\Delta v_i^*$ / $cm^3 \cdot mol^{-1}$ | $V_i^*$ / $cm^3 \cdot mol^{-1}$ | $P_i^*$ / $J \cdot cm^{-3}$ |
|---|---|---|---|---|---|---|---|---|
| DPA | 138.00[b] | 1.24[b] | 1217[b] | 0.55[c] | − 7.5[c] | − 2.8[c] | 107.58 | 470.5 |
| DBA | 171.06[b] | 1.09[b] | 1059[b] | 0.16[c] | − 6.5[c] | − 3.4[c] | 135.59 | 481 |
| BA | 99.87[b] | 1.30[b] | 1149[b] | 0.96[d] | − 13.2[d] | − 2.8[d] | 77.65 | 515.8 |
| HxA | 133.11[b] | 1.12[b] | 971[b] | 0.78[e] | − 13.2[e] | − 2.8[e] | 105.99 | 506.2 |
| Aniline | 91.53[f] | 0.85[f] | 468[f] | 14.8[f] | − 15[f] | − 12[f] | 79.82 | 541.6 |
| DMF | 77.42[g] | 1.01[g] | 659.4[g] | | | | 61.97 | 711.4 |
| DMA | 93.05[b] | 0.98[b] | 653.5[b] | | | | 74.82 | 691.4 |
| NMA | 76.94[h] | 0.869[h] | 612[h] | 72.5[h] | − 25[h] | − 3.6[h] | 63.95 | 561.6 |

[a] $V_{mi}$, molar volume; $\alpha_{Pi}$, isobaric thermal expansion coefficient; $\kappa_{Ti}$, isothermal compressibility; $K_i$, equilibrium constant,; $V_i^*$ and $P_i^*$, reduction parameters for volume and pressure, respectively; $\Delta h_i^*$, hydrogen bonding enthalpy; $\Delta v_i^*$, self-association volume; [b][s9]; [c][s10]; [d][s11]; [e][s12]; [f][s13]; [g][s8]; [h][s14] ($T$ = 303.15 K)

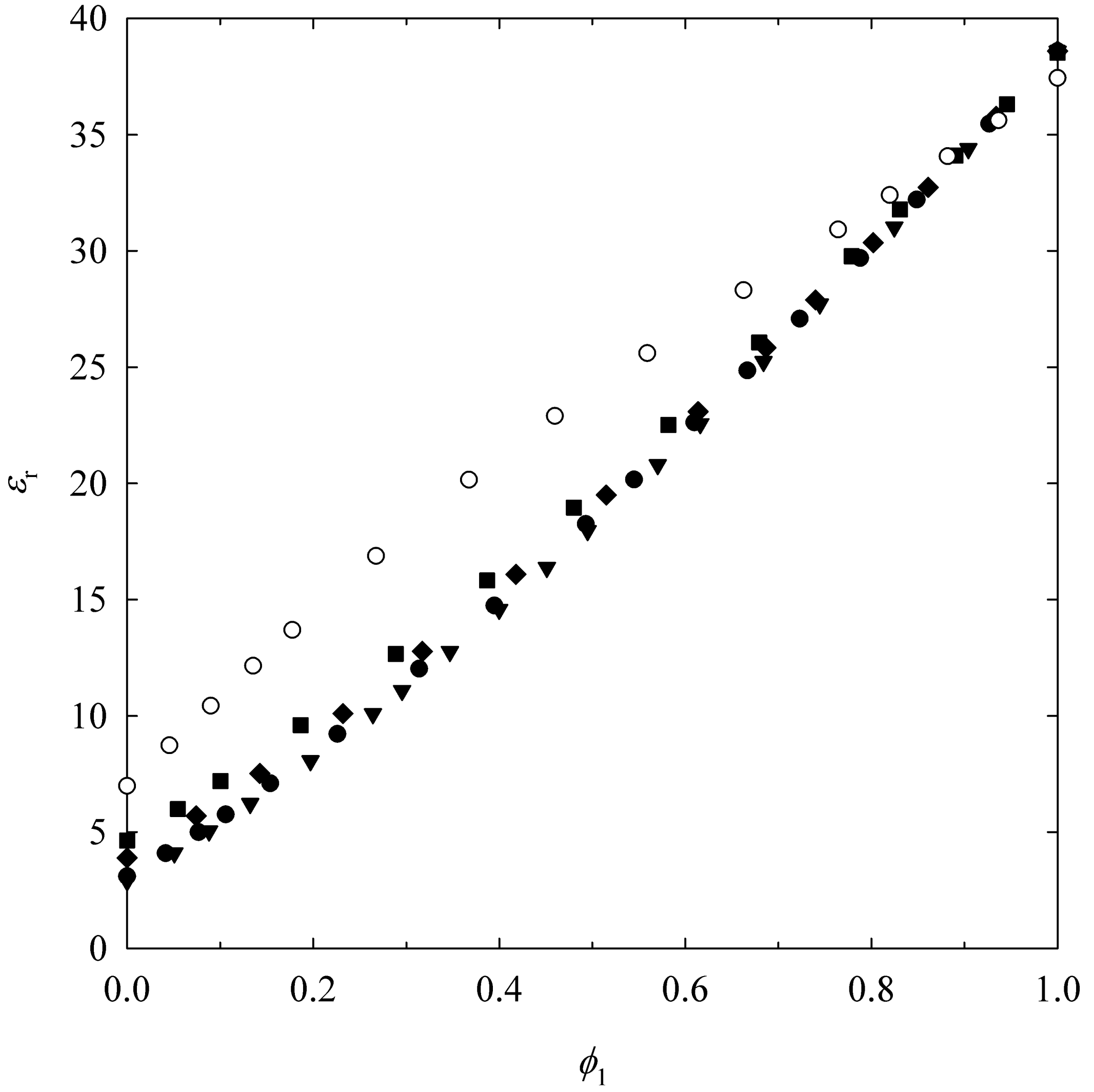

Figure S1

Relative permittivities, $\varepsilon_r$, of amide (1) + amine (2) systems at 0.1 MPa; 298.15 K and 1 MHz. (●), DMA + DPA; (▼), DMA + DBA; (■), DMA + BA; (◆), DMA + HxA; (○), DMF + aniline.

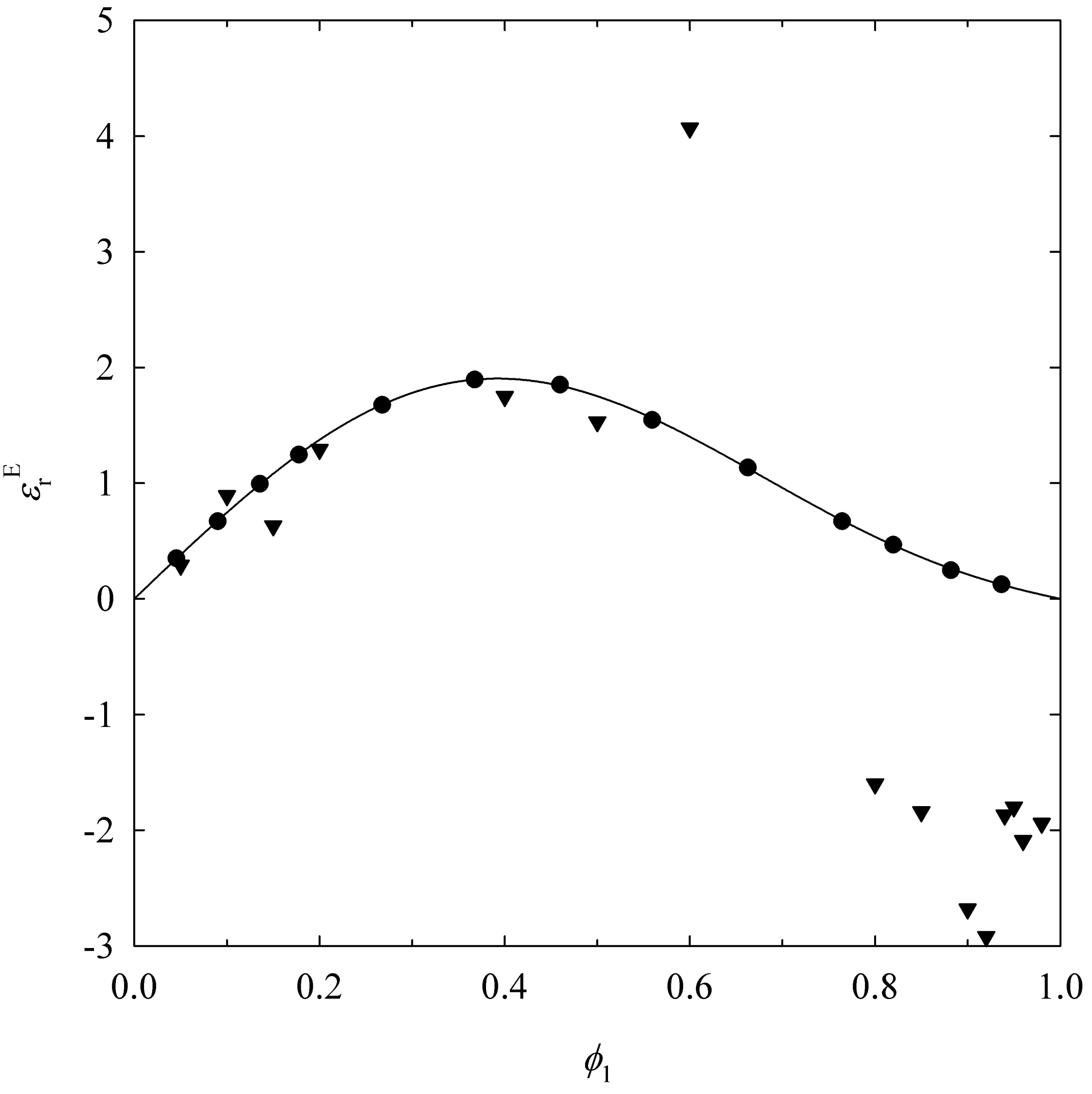

Figure S2

Excess relative permittivities, $\varepsilon_r^E$, for the DMF (1) + aniline (2) system at 0.1 MPa and 303.15 K. (●), this work; (▼), [s15]. Solid line, calculations with equation (3) using the coefficients from Table 4.

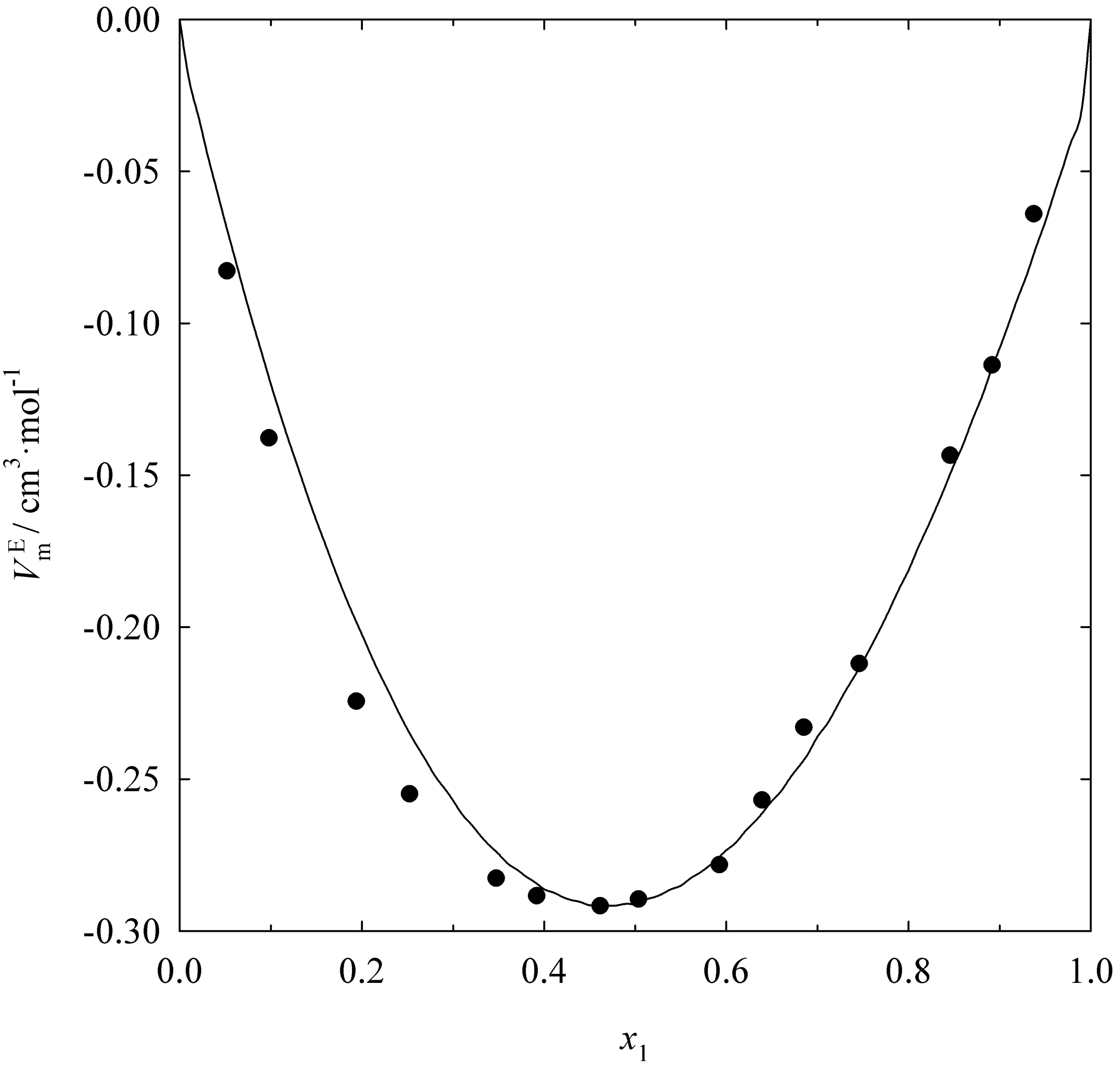

Figure S3

Excess molar volume, $V_m^E$ for the DPA (1) + DMF (2) mixture at 298.15 K and 0.1 MPa. Points experimental results [s8]. Solid line, ERAS calculations with parameters from Table 5.

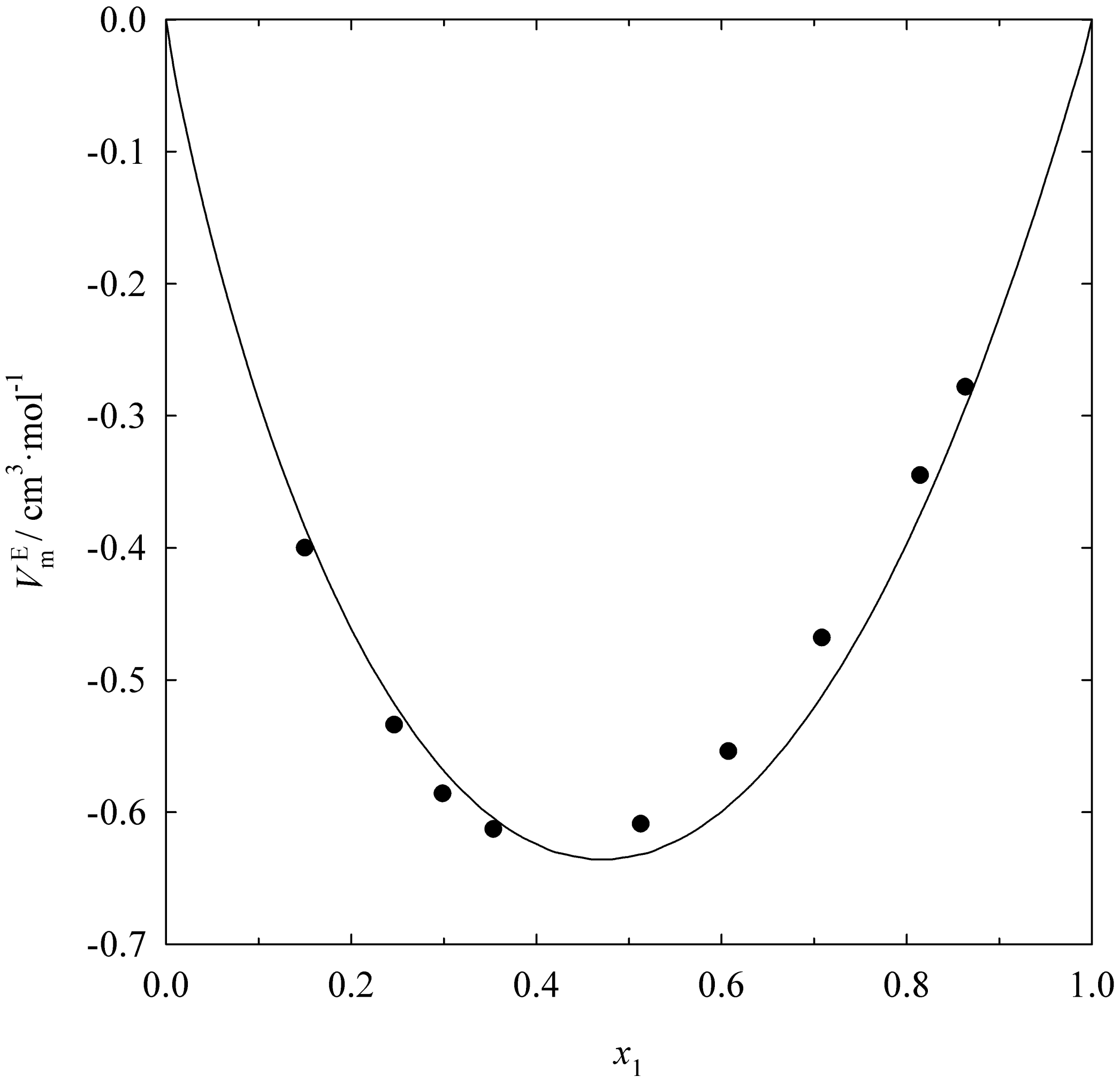

Figure S4

Excess molar volume, $V_m^E$ for the aniline (1) + DMA (2) mixture at 303.15 K and 0.1 MPa. Points experimental results [s16]. Solid line, ERAS calculations with parameters from Table 5.

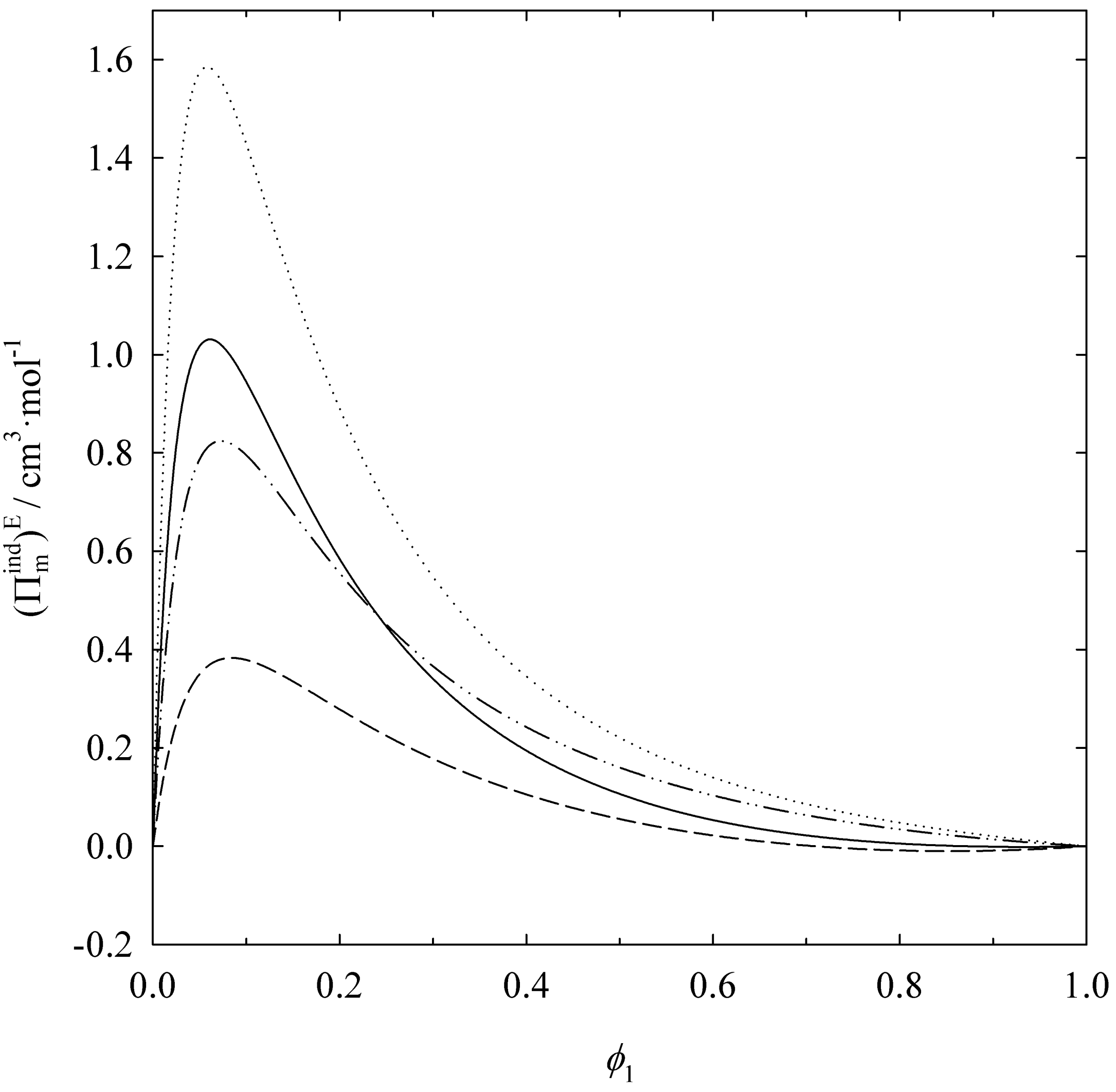

Figure S5

Excess molar induced polarizability, $(\Pi_m^{ind})^E$, for DMA (1) + linear amine (2) systems at 0.1 MPa and 298.15 K. (——), DPA; (···), DBA; (– – –), BA; (··–··), HxA.

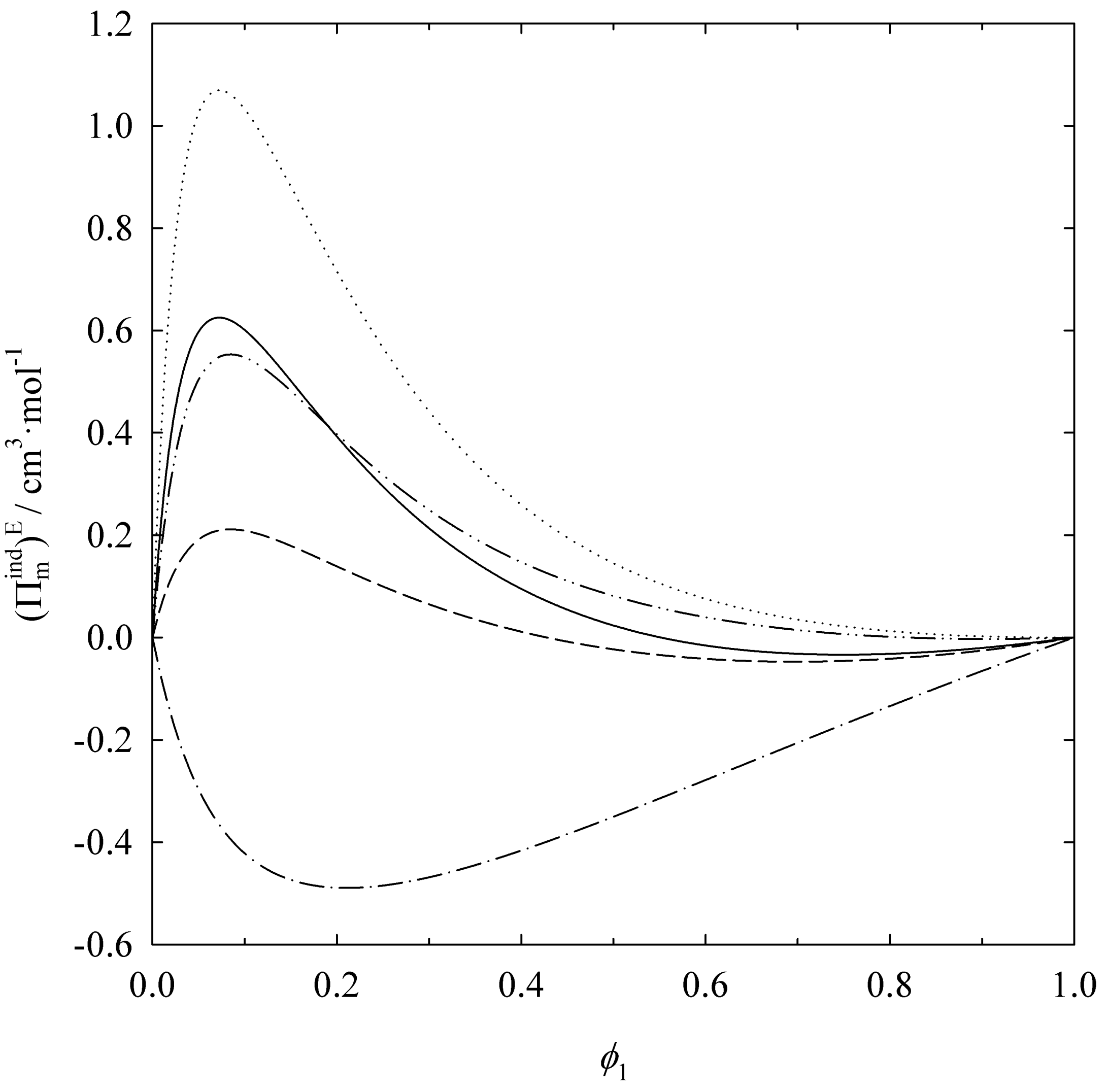

Figure S6

Excess molar induced polarizability, $(\Pi_m^{ind})^E$, for DMF (1) + linear amine (2) [s17], or + aniline (this work) systems at 0.1 MPa and 298.15 K. (——), DPA; (···), DBA; (– – –), BA; (··–··), HxA; (·–·), aniline.